\pdfoutput=1
\RequirePackage{etoolbox}
\documentclass[
	a4paper,
	rgb,dvipsnames,x11names, 
	]{bmcart}[2014/01/24]
\usepackage{amssymb}
\usepackage{amsfonts}
\usepackage[cmex10]{amsmath}
\usepackage{amsxtra}
\usepackage{scalerel}
\usepackage{mathrsfs}
\usepackage[shortlabels]{enumitem}
\usepackage{graphicx}
\usepackage[authoryear]{natbib}
\usepackage{enotez}
\usepackage{hyperref}
\usepackage{bookmark}
\usepackage{embedfile}
\usepackage{datetime}
\usepackage[utf8]{inputenc}
\usepackage{filecontents}

\newcommand{\now}{\protect\formatdate{29}{04}{2021}}

\DeclareGraphicsExtensions{.pdf}

\setenotez{backref}
\DeclareTranslation{English}{enotez-title}{Endnotes}
\DeclareInstance{enotez-list}{OEMSSOUSN}{paragraph}{
	heading   = \section*{\label{OEMSSOUSN/endnotes}#1},
	notes-sep = 0pt,
	format    = \indent\normalfont,
	number    = \textsuperscript{#1}
	}
\providecommand{\phantomsection}{}

\providecommand{\onlinedoi}[1]{\href{https://doi.org/\detokenize{#1}}{\texttt{https://doi.org/#1}}}

\makeatletter
\setattribute{additionalfiles}{skip}{\smallskipamount}
\setattribute{additionalfiles}{enumrightmargin}{12.75\p@}
\newbox\additionalfiles@box
\newlength\additionalfilelabelwidth
\newcommand\additionalfilesname{Additional files}
\newcommand\additionalfilename{Additional file}
\newenvironment{additionalfiles}[1][10.1109/5.771073]{%
	\section*{\additionalfilesname}%
	\backmatter@style\backmatter@size\let\section\heading%
	\settowidth{\additionalfilelabelwidth}{\additionalfilename 9:}%
	{\bfseries \additionalfilesname} accompanies this paper at \onlinedoi{#1}.%
	\setbox\additionalfiles@box=\vbox\bgroup%
	\begin{enumerate}[%
		align=left,
		leftmargin=*,
		rightmargin=\additionalfiles@enumrightmargin,
		labelwidth={\additionalfilelabelwidth},
		label={\bfseries{{\additionalfilename} \arabic*:}},
		ref={\arabic*},
		]}%
  {\end{enumerate}%
	\egroup%
	\vskip\additionalfiles@skip
	\setlength{\fboxsep}{5\p@}%
	\setlength{\fboxrule}{0.5\p@}%
	\fcolorbox{bmcblue}{white}{\box\additionalfiles@box}%
	}
\newcommand{\embedadditionalfile}[2][]{\IfFileExists{#2}{\embedfile[#1]{#2}}{}}
\makeatother

\startlocaldefs
\newlength{\halfheight}
\newcommand{\oneonehalf}{\scaleto{\frac{1}{2}}{0.9\halfheight}}
\newcommand{\usnStrSeg}[1]{\bar{s}_{#1}}
\newcommand{\usnDefAng}[2]{\ifx#2\empty\delta_{#1}\else\delta_{#1,#2}\fi}
\newcommand{\symess}{\mathfrak{s}}
\DeclareMathOperator{\stSu}{Su}
\DeclareMathOperator{\stSD}{SD}
\DeclareMathOperator{\mefPr}{\mathnormal{p}}
\DeclareMathOperator{\mefZ}{\mathnormal{Z}}
\DeclareMathOperator{\meffun}{\mathnormal{f}}
\DeclareMathOperator{\mefPhi}{\Phi}
\newcommand{\siftrn}[1]{\widetilde{#1}}
\newcommand{\sotaStrStp}[2][\empty]{\ifx#1\empty\mathrm{s}_{#2}\else\mathrm{s}^{#1}_{#2}\fi}
\endlocaldefs

\renewcommand\additionalfilesname{Supplementary information}
\renewcommand{\doi}[1]{\href{https://doi.org/\detokenize{#1}}{\texttt{https://doi.org/#1}}}

\begin{document}

\begin{frontmatter}

\begin{fmbox}
\dochead{Research\hfill\now}

\title{On equilibrium Metropolis simulations\\ on self-organized urban street networks}

\author[
	addressref={NYUAD,RONININSTITUTE},
	corref={RONININSTITUTE},
	email={jerome.benoit@ronininstitute.org}
	]{\inits{JGMB}\fnm{J{\'e}r{\^o}me GM} \snm{Benoit}} 
\author[
	addressref={NYUAD,NYUTSE},
	email={sej7@nyu.edu}
	]{\inits{SEGJ}\fnm{Saif Eddin G} \snm{Jabari}} 

\address[id=NYUAD]{%
	\orgname{New York University Abu~Dhabi},
	\street{Saadiyat Island},
	\postcode{POB 129188},
	\city{Abu~Dhabi},
	\cny{UAE}
	}
\address[id=RONININSTITUTE]{%
	\orgname{Ronin Institute},
	\postcode{NJ 07043},
	\city{Montclair},
	\cny{USA}
	}
\address[id=NYUTSE]{%
	\orgname{New York University Tandon~School~of~Engineering},
	\street{Brooklyn},
	\postcode{NY 11201},
	\city{New~York},
	\cny{USA}
	}

\end{fmbox}

\begin{abstractbox}

\begin{abstract}
Urban street networks of unplanned or self-organized cities
typically exhibit astonishing scale-free patterns.
This scale-freeness can be shown,
within the maximum entropy formalism (\textnormal{MaxEnt}),
as the manifestation of a fluctuating system that preserves on average some amount of information.
Monte Carlo methods that can further this perspective are cruelly missing.
Here we adapt to self-organized urban street networks the Metropolis algorithm.
The \textit{``coming to equilibrium''} distribution is established with
\textnormal{MaxEnt}
by taking scale-freeness as prior hypothesis
along with symmetry-conservation arguments.
The equilibrium parameter is the scaling;
its concomitant extensive quantity is,
assuming our lack of knowledge,
an amount of information.
To design an ergodic dynamics,
we disentangle the state-of-the-art street generating paradigms based on non\-overlapping walks
into layout-at-junction dynamics.
Our adaptation
reminisces the
single-spin-flip Metropolis algorithm for Ising models.
We thus expect Metropolis simulations to reveal that
self-organized urban street networks,
besides sustaining scale-freeness over a wide range of scalings,
undergo a crossover as scaling varies
---
literature argues for a small-world crossover.
Simulations for Central London
are consistent
against
the state-of-the-art outputs
over a realistic range of scaling exponents.
Our illustrative Watts-Strogatz phase diagram with scaling as rewiring parameter
demonstrates a small-world crossover curving within the realistic window $2$--$3$;
it also shows that the state-of-the-art outputs underlie relatively large worlds.
Our Metropolis adaptation to self-organized urban street networks
thusly appears as a scaling variant of the Watts-Strogatz model.
Such insights may ultimately allow the urban profession
to anticipate self-organization
or unplanned evolution
of urban street networks.
\end{abstract}

\begin{keyword}
\kwd{Urban street networks}
\kwd{Self-organization}
\kwd{Scale-freeness}
\kwd{Metropolis algorithm}
\kwd{MaxEnt}
\kwd{Symmetries}
\kwd{Conserved quantities}
\kwd{Self-similarity}
\kwd{Surprisal}
\kwd{Graph matchings}
\kwd{Ising model}
\kwd{Watts-Strogatz model}
\kwd{Small-world crossover}
\end{keyword}

\end{abstractbox}

\end{frontmatter}

\hypersetup{%
	pdfdisplaydoctitle=true,
	pdftitle={On equilibrium Metropolis simulations on self-organized urban street networks},
	pdfauthor={J\'er\^ome~Benoit (ORCID: 0000-0003-1226-6757) and Saif~Eddin~Jabari (ORCID: 0000-0002-2314-5312)},
	pdfsubject={%
		Applied Network Science special issue:
		ComplexNetwork 2019
		(%
			The 8th International Conference on Complex Networks and Their Applications
			- Lisbon 10-12 December, 2019, (Portugal)%
			)
		},
	pdfkeywords={
		Urban street networks,
		[Spatial networks],
		Self-organization,
		Scale-freeness,
		Metropolis algorithm,
		MaxEnt,
		Symmetries,
		Conserved quantities,
		Self-similarity,
		Surprisal,
		[Discrete Pareto distribution],
		Graph matchings,
		[Hosoya index],
		Ising model,
		Watts-Strogatz model,
		Small-world crossover;
		Interdisciplinary physics,
		Complex System,
		Statistical Physics.%
		},
	pdfcreator={\LaTeXe{} and its friends (git-id: oemssousn-1.3.0-0-g14b62c9)},
	pdfpagelayout=SinglePage,
	pdfpagemode=UseOutlines,
	pdfstartpage=1,
	pdfhighlight=/O,
	pdfview=FitH,
	pdfstartview=FitH,
	hyperfootnotes=true,
	colorlinks=true,
	allcolors=RedOrange,
	citecolor=RoyalBlue3,
	urlcolor=RoyalBlue3,
	linkcolor=RoyalBlue3,
	bookmarksnumbered=true,
	bookmarksopen=true,
	bookmarksopenlevel=3,
	}

\section*{\addcontentsline{toc}{section}{Introduction}\label{sec/introduction}Introduction}

Unplanned or self-organized cities
spontaneously
undergo scaling coherences for which a comprehensive explanation is lacking \citep{DRybskiUSL2019}.
Scaling coherence,
or scale-freeness,
expresses apparent invariance under zooming-in or -out transformations.
The scaling coherence of the spatial organization of a city is reflected in its streets:
the streets of a self-organized city typically follow a scale-free behaviour
which has attracted much attention from observational and theoretical researchers
\citep{MRosvall2005,PortaTNAUSDA2006,CrucittiCMSNUS2006,BJiangSZhaoJYin2008}.
We recently linked the scale-freeness of self-organized urban street networks
to a preservation principle through a fluctuating mesoscopic model \citep{JBenoitOPSOUSN2019,SESOPLUSN}.

The invoked
preservation principle is the Jaynes's Maximum Entropy principle \citep{ETJaynes1957I,ETJaynesPTLS,ALawrencePP2019}.
This principle assesses
the most plausible probability distribution of a fluctuating system
according to moment constraints.
We inversely applied it by envisioning streets
as mesoscopic objects governed by social interactions \citep{JBenoitOPSOUSN2019,SESOPLUSN}.
We reflect the scaling coherence by randomly distributing their numbers of configurations
according to a scale-free distribution,
specifically,
a discrete Pareto distribution \citep{AClausetCRShaliziMEJNewman2009}.
The discrete Pareto distribution
results from a constraint on the first logarithm moment \citep{YDover2004}.
Since their configurations are equally probable due to our lack of knowledge,
this constraint interprets itself as an information measure preservation.
The predominance of a number of vital connections among social connections asymptotically leads to
a discrete Pareto distribution for the number of junctions per street.
We have what is observed among self-organized urban street networks.
However promising the approach appears,
we need to investigate it completely with some specific tools.

To study such fluctuating models,
analytical and simulational methods are usually employed as complementary methods
to obtain more complete and accurate interpretations.
Our analytical framework is the \emph{maximum entropy formalism},
a general formalism of modern probability theory
partially inherited from statistical physics \citep{WTGrandyJrFSMET,ETJaynesPTLS,ALawrencePP2019}.
For simulating fluctuating systems,
physicists mostly rely on random sampling algorithms based on Markov chain Monte Carlo methods,
often abbreviated as \emph{Monte Carlo methods} \citep{MCMSP,DPLandauKBinderGMCSSP}.
Each thus-generated random sample enables us to obtain numerical results that
we can confront to theoretical ones.
The Monte Carlo method of first choice remains
the algorithm pioneered
by Nicolas Metropolis
and his co-workers
\citep{NMetropolis1953,MCMSP}.

\begin{figure}[hbtp]
	\begin{center}
		\includegraphics[width=0.95\linewidth]{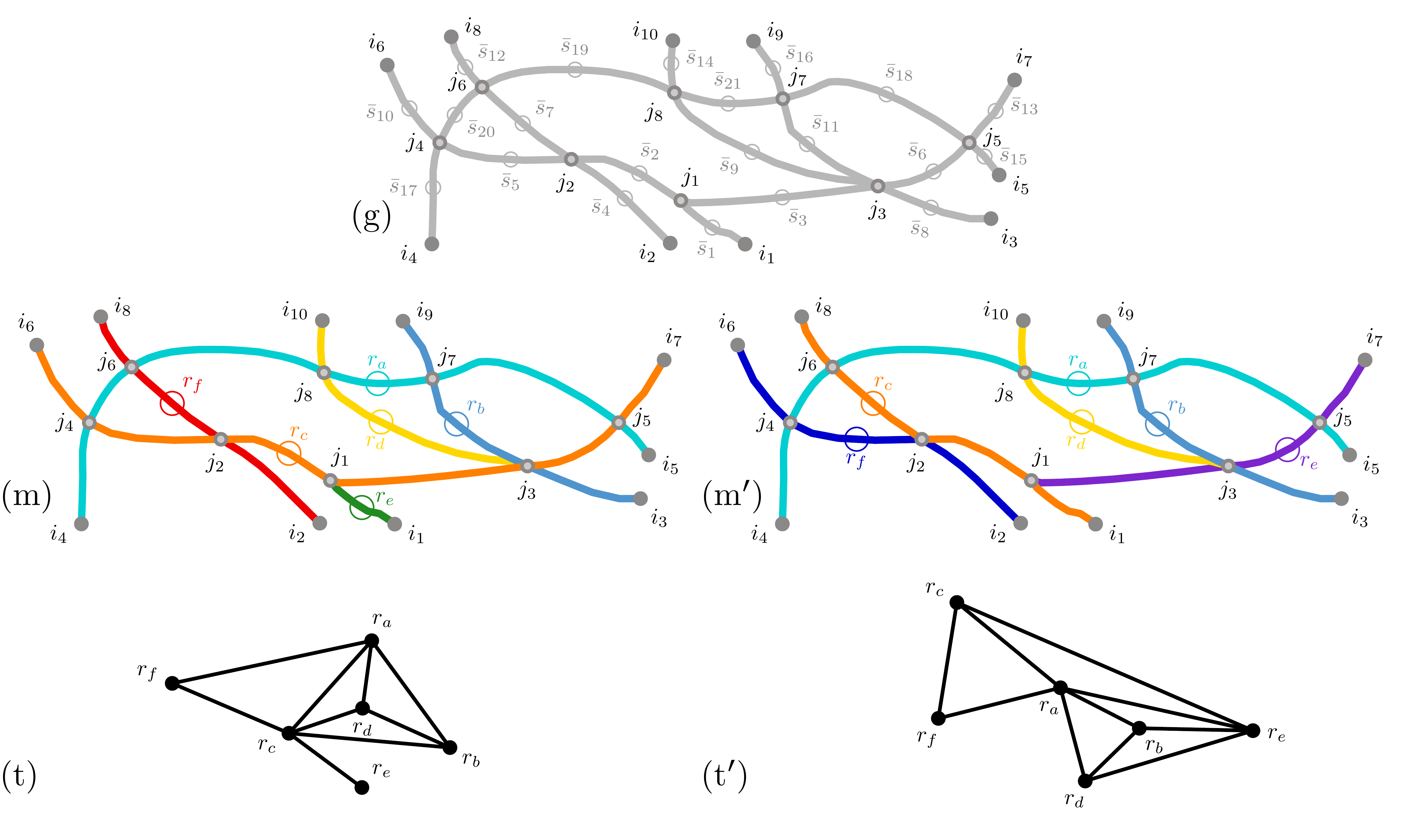}
	\end{center}
	\caption{\label{OEMSSOUSN/fig/introduction/notionalexamples}%
		Notional urban street network meant to pattern
		throughout this paper
		a real-world urban street network.
		The planar graph representation ($\mathrm{g}$)
		emphasizes a literal geometric interpretation
		where
		junctions $j_{\ast}$ (and impasses $i_{\ast}$) are nodes
		and
		street-segments $\usnStrSeg{\ast}$ are edges.
		The street maps ($\mathrm{m}$) and ($\mathrm{m}^{\prime}$) show
		two of the possible configurations of streets associable to graph ($\mathrm{g}$).
		The information networks ($\mathrm{t}$) and ($\mathrm{t}^{\prime}$) emphasize
		the topological information contained
		in street maps ($\mathrm{m}$) and ($\mathrm{m}^{\prime}$),
		respectively:
		they associate streets $r_{\ast}$ to nodes
		and they link streets sharing common junctions $j_{\ast}$.
		Information networks
		of self-organized or unplanned urban street networks
		exhibit
		in general
		a scale-free valence distribution,
		namely,
		they are scale-free networks.
		This observational fact has led us to
		a fluctuating model
		for which configurations of streets are fluctuating
		as part of a social process.
		So,
		along this paper,
		a street map such as ($\mathrm{m}$) or ($\mathrm{m}^{\prime}$)
		is abstracted as a state\endnotemark[\ref{OEMSSOUSN/en/state}]
		of a fluctuating system.
		}
\end{figure}

Strictly speaking
the \emph{Metropolis algorithm} may apply
to
configurations of streets or their associated
information networks.
An information network  \citep{MRosvall2005,PortaTNAUSDA2006} is
a dual network representation of an urban street network that
(i) associates each street to a node,
and
(ii) links each pair of nodes
(streets)
sharing a common junction
(see Figure~\ref{OEMSSOUSN/fig/introduction/notionalexamples} for illustration).
It is this dual graph representation that
reveals the underlying scale-freeness \citep{PortaTNAUSDA2006,CrucittiCMSNUS2006,BJiangSZhaoJYin2008}.
For instance,
the valence distribution of an information network
associated to a self-organized urban street network
typically follows a discrete Pareto distribution \citep{AClausetCRShaliziMEJNewman2009}.
This observed scale-freeness provides
a clue to find the prior hypothesis \citep{WTGrandyJrFSMET,ETJaynesPTLS}
necessary to construct a fluctuating mesoscopic model for the streets,
that is,
to model
the probability distribution to which Monte Carlo simulations
are \emph{``coming to equilibrium''} \citep{MCMSP,DPLandauKBinderGMCSSP}.
For mimicking fluctuating transitions,
we may use the property that
one information network transforms into another
when a junction alters its street layout.

Basically,
a Monte Carlo simulation
iterates a Markov process
for generating a Markov chain of states,
a sequence of states whose every state depends only on its predecessor \citep{MCMSP,DPLandauKBinderGMCSSP}.%
\endnote{%
A \emph{state} is a set of quantities
completely describing a system
which does not include anything about its history.
Along this notion,
a \emph{dynamics} is a map
associating to a state another state
which does not depend on the past states.
A Markov process is a dynamics.%
}\phantomsection\label{OEMSSOUSN/en/state} 
Here a state is
any configuration of streets or its associated information network
(see Figure~\ref{OEMSSOUSN/fig/introduction/notionalexamples} for illustration).
The Markov process is built so that the Markov chain reaches,
when it is iterated enough times
starting from any arbitrary state,
the prescribed statistical equilibrium.
To achieve this,
the Markov process has to fulfil
(i) the \emph{condition of detailed balance}
and
(ii) the \emph{condition of ergodicity}.
The Metropolis algorithm is essentially an implementation choice for the former.
The implementation of the condition of ergodicity relies on the details of the systems.
The objective of this work is twofold.
First,
to present
how a Metropolis algorithm adaptation can compel these two conditions for self-organized urban street networks.
Second,
to apprehend whether or not Metropolis simulations can provide
pertinent \textit{``experimental''} data to investigate their scaling coherence.

The rest of the paper presents our
innovative modeling approach
as follows.
The second section
carries out the two requested conditions.
Firstly,
once the probability distribution to come to equilibrium is established,
the condition of detailed balance reduces to writing down
the emblematic Metropolis acceptance ratio.
Secondly,
a short analysis enables us to disentangle
the state-of-the-art paradigms for generating information networks
into a constrained ergodic dynamics,
which nonetheless recalls the classical single-spin-flip ergodic dynamics.
This dynamics can potentially become unconstrained.
Eventually,
our Metropolis adaptation implements itself
and compares easily against the classical single-spin-flip adaptation for Ising models.
Next,
the third section compares,
over a wide range of scaling exponents,
Metropolis generation series
against state-of-the-art outputs
for Central London (United Kingdom).
The range of consistency renders
scaling investigations around their accepted scaling values feasible.
As illustration,
we plot the Watts-Strogatz phase diagram
with scaling as rewiring parameter.
We demonstrate thusly a small-world crossover
curving at realistic scaling values.
Accordingly the state-of-the-art outputs underlie relatively large worlds.
In the concluding section,
after a summary of the findings,
we point how the presented
methodology
may contribute,
as part of a fluctuating system approach,
to change
our perspective
on urban street networks and,
by extension,
on cities.

\section*{\addcontentsline{toc}{section}{Implementation of the Metropolis algorithm}\label{sec/modelization}Implementation of the Metropolis algorithm}

This section shows how we can apply
the classical Metropolis algorithm
on
unplanned or
self-organized urban street networks
to generate scale-free streets.
We first adapt the most emblematic part,
then
we design
two appropriate dynamics.
Each dynamics aims to create from any current configuration of streets
a new
one.
The emblematic part tells us whether or not to accept
the new configurations of streets
in order for their sequences to tend to a prescribed statistical equilibrium.

\subsection*{\addcontentsline{toc}{subsection}{The emblematic Metropolis acceptance ratio}\label{sec/modelization/acceptanceratio}The emblematic Metropolis acceptance ratio}

Typically Monte Carlo methods are applied to thermal systems.
So applying them to a non-thermal system requires
the extra preliminary work to frame the statistics of its steady fluctuations.
The framework provided by the maximum entropy formalism
allows us to derive
an equilibrium distribution
which is relevant to our scale-free system.
This first achievement of our paper is necessary to implement any Monte Carlo method.
The resulting Metropolis acceptance ratio takes a typical form.

\subsubsection*{\addcontentsline{toc}{subsubsection}{Scale-freeness as available information}\label{sec/modelization/acceptanceratio/scalefreeness}Scale-freeness as available information}

In the classic literature,
the prescribed equilibrium distribution is
de facto
the Boltzmann distribution \citep{MCMSP,DPLandauKBinderGMCSSP}.
The same modern tools that derive the Boltzmann distribution from a conservation argument
allows us to establish
the prescribed equilibrium distribution of a scale-free system
through a symmetry argument.
We obtain a discrete Pareto distribution of an undefined quantity.
This result should be folklore in some area,
but we could not locate it in the literature.

At thermal equilibrium,
the probability $p_{\mu}$ for a thermal system to occupy
any state\endnotemark[\ref{OEMSSOUSN/en/state}]
$\mu$ is assumed to yield
the Boltzmann distribution
\begin{equation}\label{OEMSSOUSN/eq/MCM/energy/probability}
	p_{\mu}\propto\mathrm{e}^{-\beta{E_{\mu}}}
\end{equation}
with $E_{\mu}$ the energy of state $\mu$
and $\beta$ the inverse temperature \citep{MCMSP,DPLandauKBinderGMCSSP,WTGrandyJrFSMET}.
We have $\beta=1/kT$ with $k$ the Boltzmann constant
and $T$ the temperature.
Nowadays
the probability distribution \eqref{OEMSSOUSN/eq/MCM/energy/probability}
can easily be derived by applying
the \emph{principle of maximum entropy}
(\textsc{MaxEnt})
formulated
by \citet{ETJaynes1957I}
as a general principle of probability theory \citep{ETJaynesPTLS,WTGrandyJrFSMET,ALawrencePP2019}.
Within the maximum entropy formalism,
Boltzmann probability \eqref{OEMSSOUSN/eq/MCM/energy/probability} becomes
the most plausible probability distribution that
preserves the total energy of the system on average.
This preservation is formally a constraint imposed on the mean of the energy.
In practice
the constraint is treated as a standard variational problem \citep{ETJaynesPTLS,WTGrandyJrFSMET}
using the method of Lagrangian multipliers \citep[see for example][App.~{2}]{DApplebaumPIIA}.
The Lagrangian writes \citep{ETJaynesPTLS}
\begin{equation}\label{OEMSSOUSN/eq/MCM/Lagrangian/energy/expression}
	{\mathcal{L}_{\mathrm{TE}}}\left(\left\{p_{\mu}\right\};\nu,\beta\right) =
		-\sum_{\mu} p_{\mu}\ln{p_{\mu}}
		-\nu\biggl[
				\sum_{\mu} p_{\mu} - 1
				\biggr]
		-\beta\biggl[
				\sum_{\mu} p_{\mu} E_{\mu} - {\left\langle{E}\right\rangle}
				\biggr]
\end{equation}
where
$p_{\mu}$ is our unknown probability distribution
and the first Lagrange multiplier~$\nu$ forces its normalization,
while
the second Lagrange multiplier $\beta$ imposes
the mean energy to have the constant energy value ${\left\langle{E}\right\rangle}$.
The stationary solution of Lagrangian \eqref{OEMSSOUSN/eq/MCM/Lagrangian/energy/expression} is
the desired
probability distribution $p_{\mu}$ \citep{ETJaynesPTLS};
we have
\begin{equation}\label{OEMSSOUSN/eq/MCM/Lagrangian/energy/variation}
	0 = \delta {\mathcal{L}_{\mathrm{TE}}}\left(\left\{p_{\mu}\right\};\nu,\beta\right)
		= \sum_{\mu} \Bigl[
			-\ln{p_{\mu}}
			-\left(\nu+1\right)
			-\beta E_{\mu}
			\Bigr]
		\,
		\delta{p_{\mu}}
\end{equation}
for arbitrarily small variations $\delta{p_{\mu}}$ of $p_{\mu}$.
Resolving \eqref{OEMSSOUSN/eq/MCM/Lagrangian/energy/variation} immediately gives
\begin{equation}\label{OEMSSOUSN/eq/MCM/Lagrangian/energy/solution}
	p_{\mu} =
		\frac{\mathrm{e}^{-\beta{E_{\mu}}}}{\mefZ(\beta)}
	\qquad\text{with}\qquad
	{\mefZ(\beta)} = \sum_{\mu} \mathrm{e}^{-\beta{E_{\mu}}}
\end{equation}
as \emph{partition function} \citep{ETJaynesPTLS,WTGrandyJrFSMET}.
Probability distribution \eqref{OEMSSOUSN/eq/MCM/Lagrangian/energy/solution}
is Boltzmann probability \eqref{OEMSSOUSN/eq/MCM/energy/probability}
expressed
in its \emph{canonical form} \citep{ETJaynesPTLS,WTGrandyJrFSMET}.
If the maximum entropy formalism tells us how to treat total energy preservation,
noticeably
it does not tell us why we choose this constraint over another.
Formally the preservation of the total energy
is part of the initial \emph{hypothesis} or \emph{available information} \citep{ETJaynesPTLS,WTGrandyJrFSMET}
that we have on systems in thermal equilibrium.

For self-organized urban street networks,
our only available information is scale-freeness.
However scale-freeness is not a preserved quantity but rather a property \citep{HEStanleyIPTCP}.
But,
at the same time,
scale-freeness of a self-organized information network
may result from a self-similarity inherited
from its self-organized city \citep{AClausetSIRN2006,MBatty2008}.
Self-similarity is a symmetry \citep{BMandelbrot1982},
a transformation that lets an object or a system stay invariant.
Symmetries play a fundamental role in modern physics \citep{DJGrossTRSFP1996,DRomeroMaltrana2015,MASerrano2018}.
A general consensus
in physics
is that
an invariance to a transformation underlies a preserved entity,
and vice versa \citep{DJGrossTRSFP1996,DRomeroMaltrana2015}.
Let us see how this idea applies here.
For our purpose,
we must first rewrite Lagrangian \eqref{OEMSSOUSN/eq/MCM/Lagrangian/energy/expression}
in the more generic form
\begin{equation}\label{OEMSSOUSN/eq/MCM/Lagrangian/scaling/preexpression}
	{\mathcal{L}_{\meffun}}\left(\left\{p_{\mu}\right\};\nu,\lambda\right) =
		-\sum_{\mu} p_{\mu}\ln{p_{\mu}}
		-\nu\biggl[
				\sum_{\mu} p_{\mu} - 1
				\biggr]
		-\lambda\biggl[
				\sum_{\mu} p_{\mu} \meffun(X_{\mu}) - {\left\langle{\meffun({X})}\right\rangle}
				\biggr]
\end{equation}
where $X$ is an extensive quantity whose each value $X_{\mu}$ describes state~$\mu$.
An extensive quantity scales linearly under scaling transformations.
The new second Lagrange multiplier $\lambda$
imposes our unknown constraint which
expresses in terms of an unknown function $\meffun$ acting on $X$.
It literally coerces the mean value of ${\meffun({X})}$
to have the constant value ${\left\langle{\meffun({X})}\right\rangle}$.
For the sake of demonstration,
we will assume exact self-similarity.
Accordingly,
under the scaling transformation
\begin{equation}\label{OEMSSOUSN/eq/MCM/Lagrangian/scaling/transformation}
	x \to \siftrn{x}=\symess\,{x}
	,
\end{equation}
a self-similar
(or homogeneous)
function $\mefPhi(x)$ will transform as
\begin{equation}\label{OEMSSOUSN/eq/MCM/Lagrangian/scaling/selfsimilarity}
	\mefPhi(x) \to \siftrn{\mefPhi(x)}=\mefPhi(\symess\,{x})=\symess^{\alpha} \mefPhi(x)
\end{equation}
with $\alpha$ a scaling exponent \citep[sec.~11.1]{HEStanleyIPTCP}.
Here,
the self-similarity invariance holds
in the unknown probability distribution ${p_{\mu}=\mefPr(X_\mu)}$.
Under transformation \eqref{OEMSSOUSN/eq/MCM/Lagrangian/scaling/transformation},
$p_{\mu}$ remains unchanged as expected;
we have
\begin{equation}\label{OEMSSOUSN/eq/MCM/Lagrangian/scaling/statisticalselfsimilarity}
	p_{\mu} \to
		\siftrn{p_{\mu}} =
		\siftrn{\mefPr(X_{\mu})} =
		\frac{\mefPr(\symess\,{X_{\mu}})}{\sum_{\mu} \mefPr(\symess\,{X_{\mu}})} =
		\frac{\symess^{\alpha}\,\mefPr({X_{\mu}})}{\sum_{\mu} \symess^{\alpha} \mefPr({X_{\mu}})} =
		\mefPr(X_{\mu}) =
		p_{\mu}
	.
\end{equation}
If we demand that Lagrangian \eqref{OEMSSOUSN/eq/MCM/Lagrangian/scaling/preexpression}
stays invariant under transformation \eqref{OEMSSOUSN/eq/MCM/Lagrangian/scaling/transformation},
then
\begin{align}\label{OEMSSOUSN/eq/MCM/Lagrangian/scaling/invariance/equation/literal}
	0
		&=
			\left({{\mathcal{L}_{\meffun}}\left(\left\{p_{\mu}\right\};\nu,\lambda\right)}\right)\sptilde-
				{{\mathcal{L}_{\meffun}}\left(\left\{p_{\mu}\right\};\nu,\lambda\right)}
			\nonumber\\
		&=
			-\lambda\,\biggl\{
					\sum_{\mu} p_{\mu} \bigl[ \meffun(\symess\,{X_{\mu}}) - \meffun(X_{\mu}) \bigr]
					- \left\langle{\meffun({\symess\,X})}-{\meffun({X})}\right\rangle
					\biggr\}
			\nonumber\\
		&=
			-\lambda\,\biggl\{
					\sum_{\mu} p_{\mu} \bigl[{\meffun(\symess\,{X_{\mu}})}-{\meffun(\symess)}-{\meffun(X_{\mu})}\bigr]
					- \left\langle{\meffun({\symess\,X})}-{\meffun(\symess)}-{\meffun({X})}\right\rangle
					\biggr\}
\end{align}
for any scaling factor $\symess$
and
any possible probability distribution $p_{\mu}$.
Hence,
the unknown function $\meffun$ satisfies the functional equation
\begin{equation}\label{OEMSSOUSN/eq/MCM/Lagrangian/scaling/invariance/equation/Cauchy}
	\meffun(\symess\,{x}) = \meffun(\symess) + \meffun(x)
	.
\end{equation}
When $X$ takes only positive values $x$,
the most general solution of \eqref{OEMSSOUSN/eq/MCM/Lagrangian/scaling/invariance/equation/Cauchy}
which is continuous
is
\begin{equation}\label{OEMSSOUSN/eq/MCM/Lagrangian/scaling/invariance/solution/general}
	\meffun(x) = K \ln{x}
\end{equation}
with $K$ a constant \citep[Th.~2.1.2(2)]{JAczelLFEA1966}. 
Substituting solution \eqref{OEMSSOUSN/eq/MCM/Lagrangian/scaling/invariance/solution/general}
into the generic Lagrangian \eqref{OEMSSOUSN/eq/MCM/Lagrangian/scaling/preexpression}
gives the self-similar Lagrangian
\begin{equation}\label{OEMSSOUSN/eq/MCM/Lagrangian/scaling/expression}
	{\mathcal{L}_{\mathrm{SE}}}\left(\left\{p_{\mu}\right\};\nu,\lambda\right) =
		-\sum_{\mu} p_{\mu}\ln{p_{\mu}}
		-\nu\biggl[
				\sum_{\mu} p_{\mu} - 1
				\biggr]
		-\lambda\biggl[
				\sum_{\mu} p_{\mu} \ln{X_{\mu}} - {\left\langle{\ln{X}}\right\rangle}
				\biggr]
\end{equation}
once the useless constant $K$ is absorbed.
One easily verifies that \eqref{OEMSSOUSN/eq/MCM/Lagrangian/scaling/expression}
remains indeed unchanged
under the scaling transformation \eqref{OEMSSOUSN/eq/MCM/Lagrangian/scaling/transformation}.
The corresponding most plausible probability distribution $p_{\mu}$ yields
the stationary equation
\begin{equation}\label{OEMSSOUSN/eq/MCM/Lagrangian/scaling/variation}
	0 = \delta {\mathcal{L}_{\mathrm{SE}}}\left(\left\{p_{\mu}\right\};\nu,\lambda\right)
		= \sum_{\mu} \Bigl[
			-\ln{p_{\mu}}
			-\left(\nu+1\right)
			-\lambda \ln{X_{\mu}}
			\Bigr]
		\,
		\delta{p_{\mu}}
\end{equation}
whose solution readily writes
\begin{equation}\label{OEMSSOUSN/eq/MCM/Lagrangian/scaling/solution}
	p_{\mu} =
		\frac{{X_{\mu}}^{-\lambda}}{\mefZ(\lambda)}
	\qquad\text{with}\qquad
	{\mefZ(\lambda)} = \sum_{\mu} {X_{\mu}}^{-\lambda}
\end{equation}
in the canonical form.
One quickly checks that
probability distribution \eqref{OEMSSOUSN/eq/MCM/Lagrangian/scaling/solution} is invariant
under the scaling transformation \eqref{OEMSSOUSN/eq/MCM/Lagrangian/scaling/transformation},
as expected.
This probability distribution is known as
the discrete Pareto probability distribution \citep{AClausetCRShaliziMEJNewman2009}.
Let us summarize our result as follows.
What the maximum entropy formalism \citep{WTGrandyJrFSMET,ETJaynesPTLS,ALawrencePP2019}
combined with
the symmetry-conservation correspondence idea \citep{DJGrossTRSFP1996,DRomeroMaltrana2015}
tells us about statistically self-similar steady fluctuations is threefold:
\begin{enumerate}[i)]
	\item They follow a discrete Pareto probability distribution
		with the self-similar scaling exponent as scaling exponent.
	\item They preserve on average the logarithm of an extensive quantity.
	\item Their equilibrium parameter is the self-similar scaling exponent itself.
\end{enumerate}
However,
it can tell us nothing about the nature of the extensive quantity.

Thusly,
the prerequisite to Monte Carlo methods
for self-organized urban street networks
can be expressed as follows.
At scaling equilibrium,
the probability $p_{\mu}$
for a self-organized urban street network
to develop its streets in any state $\mu$
is assumed to yield the discrete Pareto distribution
\begin{equation}\label{OEMSSOUSN/eq/MCM/surprisal/probability/preformula}
	p_{\mu}\propto\mathrm{e}^{-\lambda\ln{X_{\mu}}}
\end{equation}
with $X_{\mu}$ the value at state $\mu$
of an extensive quantity $X$
and $\lambda$ the scaling exponent.
Still,
it remains to make
a genuine hypothesis on the extensive quantity~$X$.

\subsubsection*{\addcontentsline{toc}{subsubsection}{A surprisal-driven system}\label{sec/modelization/acceptanceratio/surprisaldriven}A surprisal-driven system}

In our context
a state $\mu$
is a possible information network,
namely
a possible configuration of streets,
that an urban street network can develop
(see Figure~\ref{OEMSSOUSN/fig/introduction/notionalexamples} for illustration).
Previous investigations show that
an information network of a self-organized urban street network typically underlies scale-freeness
\citep{PortaTNAUSDA2006,CrucittiCMSNUS2006,BJiangSZhaoJYin2008}.
Therefore,
as shown in previous section,
the distribution of their nodes
(streets)
preserves on average the logarithm of an extensive quantity,
so that this distribution is most plausibly a discrete Pareto distribution of this extensive quantity.
This extensive quantity cannot be specified due to our lack of knowledge
on information networks of self-organized urban street networks.

However,
the simplest assumption we can make is that
a self-organized urban street network is
a self-similar mesoscopic system
whose mesoscopic objects have equiprobable configurations.
Namely,
we apply to our mesoscopic objects
the \emph{principle of indifference} \citep{ETJaynesPTLS,ALawrencePP2019}.
We may call such a system
a self-similar Boltzmann-mesoscopic system.
Our extensive quantity becomes then the number of equiprobable configurations
of the mesoscopic objects.
Let us denote
by $\Pr(\Omega)$
the probability for a mesoscopic object
to have $\Omega$ possible equiprobable configurations,
and
by $o(\Omega)$
a mesoscopic object having effectively $\Omega$ possible equiprobable configurations.
With these notations,
we may say that
each mesoscopic object $o(\Omega)$ has $\Omega$ as extensive quantity.
Thence,
for each mesoscopic object $o(\Omega)$,
our extensive quantity logarithm $\ln{\Omega}$
interprets itself
either as the
Boltzmann entropy of $o(\Omega)$
or as the surprisal associated to each configuration of $o(\Omega)$.
\emph{Surprisal}
(or \emph{surprise}, or \emph{information content})
\begin{math}
	{\stSu=-\ln\circ\Pr}
\end{math}
measures
uncertainty,
astonishment,
and
knowledge
attached to
an event \citep{MTribusTT,DJCMacKayITILA,DApplebaumPIIA,JVStone2015,ALawrencePP2019}.
While the average of surprisal
over all the possible events
gives their (Shannon) entropy,
the surprisal attached to a possible event pertains its cognitive magnitude.
When an event expected to be rare occurs,
we are surprised and we feel that we learn a lot:
the larger the uncertainty before the event,
the greater the astonishment at the event,
the wider the knowledge after the event
\citep{DJCMacKayITILA,DApplebaumPIIA,ALawrencePP2019}.
And vice versa.
So that,
compared to the entropy interpretation,
the surprisal interpretation
appears in essence
finer and more cognitive.
For these reasons,
we may favour the surprisal interpretation.
The preserved moment
\begin{math}
	{\sum_{\Omega} \Pr(\Omega)\ln{\Omega}}
\end{math}
interprets
then
itself as
an amount of surprisal
that equilibria preserve on average.
We interpret thusly steady fluctuations
as a manifestation of uncertainties, astonishments, and knowledges
whose the magnitudes remain on average the same.
Presuming that this manifestation
actually
reflects a social process,
each equilibrium becomes then a match between
steady fluctuating configurations of streets
and
how city-dwellers
comprehend
their own urban street network \citep{YDover2004,JBenoitOPSOUSN2019,SESOPLUSN}.
We may expect that their comprehension reflects
their agility and proficiency
to navigate
their own urban street network
in normal or disrupted traffic.

With this assumption,
the probability $p_{\mu}$ for a self-organized urban street network
to develop an information network
(or a configuration of streets)
$\mu$ yields
\begin{equation}\label{OEMSSOUSN/eq/SOUSN/surprisal/probability/formula}
	p_{\mu}\propto\prod_{o_{\mu}\in\{s_{\mu},j_{\mu}\}}\Omega_{o_{\mu}}^{-\lambda}=\mathrm{e}^{-\lambda{S_{\mu}}}
\end{equation}
with
\begin{equation}\label{OEMSSOUSN/eq/SOUSN/surprisal/probability/def/amount/total}
	S_{\mu}=\sum_{o_{\mu}\in\{s_{\mu},j_{\mu}\}}\ln\Omega_{o_{\mu}}
\end{equation}
the total amount of surprisal for information network $\mu$;
the product (the sum) is over the streets $s_{\mu}$ and junctions $j_{\mu}$ of information network $\mu$.
Along the interpretation developed in the previous paragraph,
the total amount of surprisal $S_{\mu}$ \eqref{OEMSSOUSN/eq/SOUSN/surprisal/probability/def/amount/total}
quantifies
the comprehension of the city-dwellers for information network $\mu$.
Thus,
accordingly,
it is their comprehension that drives probability distribution \eqref{OEMSSOUSN/eq/SOUSN/surprisal/probability/formula},
that is,
the statistical equilibrium of their own urban street network.

\subsubsection*{\addcontentsline{toc}{subsubsection}{The Metropolis acceptance ratio}The Metropolis acceptance ratio}

Now that we have set up the fluctuating statistical model of our system,
we are ready to implement the emblematic part of the Metropolis algorithm.
The Metropolis algorithm holds its specificity among Monte Carlo methods
in the implementation details of the condition of detailed balance \citep{MCMSP,DPLandauKBinderGMCSSP}.
This condition assures both
that (i) each Markov chain
(or sequence)
reaches an equilibrium
and
that (ii) the equilibrium states follow the prescribed probability distribution.
It applies,
technically,
to
the probability $P(\mu\to\nu)$ of generating a state $\nu$ from a given state $\mu$
which is called the \emph{transition probability};
along the constraint
\begin{equation}
	\sum_{\nu} {P(\mu\to\nu)} = 1
	,
\end{equation}
the transition probabilities ${P(\mu\to\nu)}$ must satisfy the detailed balance equation
\begin{equation}\label{OEMSSOUSN/eq/SOUSN/MonteCarlo/detailedbalance/literal}
	p_{\mu}\,{P(\mu\to\nu)} = p_{\nu}\,{P(\nu\to\mu)}
	.
\end{equation}
Each transition probability $P(\mu\to\nu)$ may be split into two parts as
\begin{equation}\label{OEMSSOUSN/eq/SOUSN/MonteCarlo/detailedbalance/scheme}
	P(\mu\to\nu) = {g(\mu\to\nu)}\;{A(\mu\to\nu)}
	.
\end{equation}
The \emph{selection probability} $g(\mu\to\nu)$ is a probability
imposed to our algorithm for generating a new state $\nu$ given a state $\mu$,
while
the \emph{acceptance ratio} $A(\mu\to\nu)$ gives the odds of accepting
or rejecting
the move to state $\nu$ from state $\mu$.
For the Metropolis algorithm,
the selection probabilities $g(\mu\to\nu)$ for
all permitted
transitions are equal.
Scheme \eqref{OEMSSOUSN/eq/SOUSN/MonteCarlo/detailedbalance/scheme} along this choice
reduces the detailed balance equation \eqref{OEMSSOUSN/eq/SOUSN/MonteCarlo/detailedbalance/literal}
into a ratio equation
for the acceptance ratios ${A(\mu\to\nu)}$;
we have
\begin{equation}\label{OEMSSOUSN/eq/SOUSN/Metropolis/acceptanceratio/ratio/equalitychain}
	\frac{P(\mu\to\nu)}{P(\nu\to\mu)} =
		\frac{{g(\mu\to\nu)}\;{A(\mu\to\nu)}}{{g(\nu\to\mu)}\;{A(\nu\to\mu)}} =
		\frac{A(\mu\to\nu)}{A(\nu\to\mu)} =
		\frac{p_{\nu}}{p_{\mu}}
	.
\end{equation}
The last equality tells us that
the odds of accepting or rejecting
a move between two states
are in favour to the more likely of them.
This is common sense.
Nonetheless,
this still leaves open numerous possibilities.
For the Metropolis algorithm,
the more likely moves are assumed certain,
while the less likely moves get their odds adjusted
with respect to \eqref{OEMSSOUSN/eq/SOUSN/Metropolis/acceptanceratio/ratio/equalitychain};
we read
\begin{equation}
	A(\mu\to\nu)=
		\begin{cases}
			{p_{\nu}}\,{p_{\mu}}^{-1} & \text{if } {p_{\nu}}<{p_{\mu}}\\
			1 & \text{otherwise}.
		\end{cases}
\end{equation}
For our statistical model \eqref{OEMSSOUSN/eq/SOUSN/surprisal/probability/formula},
the Metropolis acceptance ratio ${A(\mu\to\nu)}$
takes the more familiar form
\begin{equation}\label{OEMSSOUSN/eq/SOUSN/Metropolis/acceptanceratio}
	A(\mu\to\nu)=
		\begin{cases}
			\mathrm{e}^{-\lambda {(S_{\nu}-S_{\mu})}} & \text{if } {S_{\nu}-S_{\mu}}>0\\
			1 & \text{otherwise}.
		\end{cases}
\end{equation}
That is to say,
if the newly selected information network $\nu$ has
a total amount of surprisal $S_{\nu}$ strictly greater than the current one $S_{\mu}$,
we accept to replace
the current information network $\mu$ by the newly selected one $\nu$
with the probability given above;
otherwise,
we accept with certainty.

\begin{figure}[hbp]
	\begin{center}
		\includegraphics[width=0.95\linewidth]{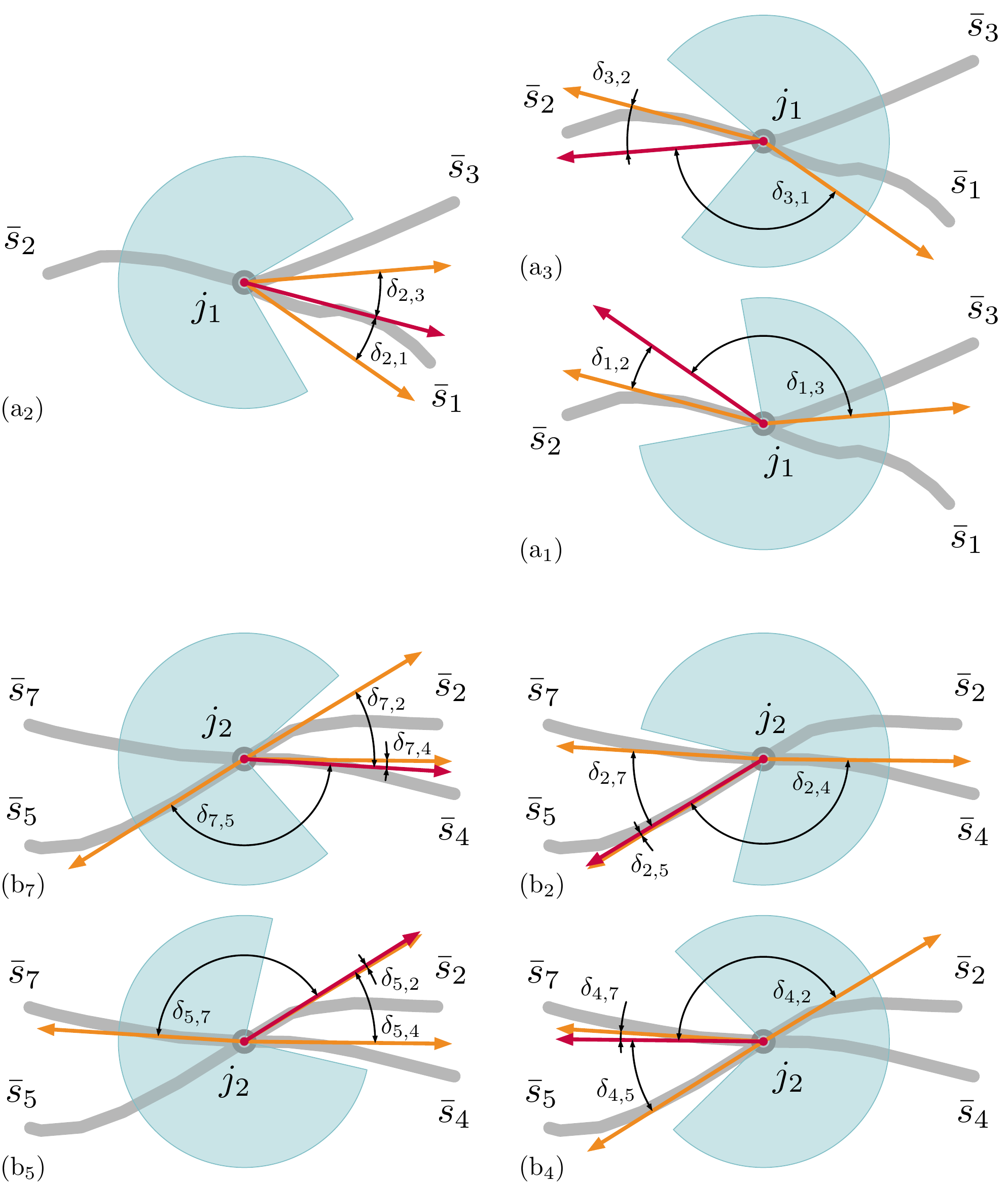}
	\end{center}
	\caption[Deflection angles at junctions]{%
			\label{OEMSSOUSN/fig/joinprinciple/deflectionangle/notionalexamples}%
		Deflection angles at junctions:
		the subfigures ($\mathrm{a}_{\star}$) and ($\mathrm{b}_{\star}$)
		show the deflection angles $\usnDefAng{\ast}{\empty}$
		for
		the junctions $j_{1}$ and $j_{2}$,
		respectively,
		from the notional example in Figure~\ref{OEMSSOUSN/fig/introduction/notionalexamples}.
		A deflection angle of a street at a junction
		is basically
		the magnitude of
		the angular change
		experienced
		at the junction
		by the tangent of the street.
		In practice,
		the street can be arbitrarily oriented
		and the deflection angle becomes
		the magnitude of the angle between the incoming and outgoing tangents.
		The transposition to pairs of street-segments at junctions is obvious.
		Each subfigure corresponds to a possible incoming street-segment.
		The subfigures actually organize
		(index)
		themselves
		according to the cardinal direction
		(index)
		of their incoming street-segment.
		For each subfigure,
		the incoming tangent
		at the junction
		is in red
		and the outgoing tangents are in orange.
		Every double-arrow arc
		between the tangents
		of an incoming street-segments $\usnStrSeg{i}$ and an outgoing one $\usnStrSeg{o}$
		indicates a deflection angle denoted by $\usnDefAng{i}{o}$
		---
		and has a radius linear with the supplementary angle $\pi-\usnDefAng{i}{o}$.
		Realistic angular changes are assumed to be bounded above.
		We have set
		the deflection angle threshold to $\pi/4$.
		The light-blue
		pie
		areas identify the forbidden deflection angles
		---
		and have the radius of any arc with the deflection angle threshold as deflection angle.
		An incoming street-segment might so continue
		its way only along
		any outgoing street-segment
		whose tangent
		or arc
		lies within the angular sector of the missing slice
		---
		and/or does not cross the pie area.
		Furthermore,
		realistic configurations of streets must obviously have no street overlap.
		No incoming street-segment
		can
		continue its way along
		an outgoing street-segment already passed through.
		The very basic idea behind the state-of-the-art for building configurations of streets
		is a loop:
		commit one choice of outgoing street-segment;
		move to its opposite junction;
		repeat.
		Figure~\ref{OEMSSOUSN/fig/streetsegments/joinedsequence/notionalexamples}
		along with
		supplementary Animation~{A\ref*{OEMSSOUSN/ani/streetsegments/joinedsequence/notionalexamples}}
		(Additional file~\ref{OEMSSOUSN/ani/streetsegments/joinedsequence/notionalexamples})
		illustrate how streets can emerge from this approach.
		In contrast,
		our approach identifies at every junction
		all the
		combinations of incoming and outgoing street-segments
		and ``flips'' them.
		Figure~\ref{OEMSSOUSN/fig/junctions/matching/maximal/notionalexamples}
		sketches why and how these combinations are actually
		maximum
		matchings,
		while
		Figure~\ref{OEMSSOUSN/fig/junctions/flipflapflopsequence/notionalexamples}
		along with
		supplementary Animation~{A\ref*{OEMSSOUSN/ani/junctions/flipflapflopsequence/notionalexamples}}
		(Additional file~\ref{OEMSSOUSN/ani/junctions/flipflapflopsequence/notionalexamples})
		illustrate a short sequence of ``flips''.
		}
\end{figure}

\subsection*{\addcontentsline{toc}{subsection}{Two simple ergodic single-junction dynamics}\label{sec/modelization/dynamics}Two simple ergodic single-junction dynamics}

The state-of-the-art
generating paradigms
are not dynamics.
This is primarily because
they build each information network from scratch.
To be a dynamics,
they should instead
create a new information network from the current one.
An analysis of their
street-oriented
paradigm
gives us clues to design relevant ergodic dynamics.
This second achievement of our paper permits us
to
concretely
adapt
the Metropolis algorithm
to self-organized urban street networks.

\subsubsection*{\addcontentsline{toc}{subsubsection}{A street is an exclusive joined sequence of street-segments}\label{sec/modelization/dynamics/streetperspective}A street is an exclusive joined sequence of street-segments}

For information networks,
nodes are streets,
basically
an exclusive sequence of successive street-segments
that are joined at junctions according to some paradigms.
By exclusive we mean that a street-segment can only belong to a single street.
This
is the perspective used in the
state-of-the-art
literature
\citep{BJiangTAUSN2004,MRosvall2005,PortaTNAUSDA2006,BJiangSZhaoJYin2008,APMsucci2014}.

An immediate paradigm is the \textit{``named street''} paradigm  \citep{BJiangTAUSN2004,BJiangSZhaoJYin2008}
which simply reproduces cadasters%
\endnote{%
A \emph{cadaster} is a comprehensive land register
maintained by either local or central authorities.
Cadasters have been used,
in some parts of the world,
for levying taxes, raising armies, setting ownerships,
etc.%
}.
Since for some cities a cadaster may not exist,
or simply reflect local habits and customs,
some studies have considered generic substitutes instead.
The choice of the paradigm may then ponder social and geographical phenomena.
A relevant parameter has appeared to be
the \emph{deflection angle} between two adjacent street-segments \citep{BJiangSZhaoJYin2008,CMolinero2017}.
Figure~\ref{OEMSSOUSN/fig/joinprinciple/deflectionangle/notionalexamples} illustrates
the notion of deflection angle
in our context
through two typical junctions.
If beyond some threshold angle any joining has to be excluded,
many possibilities remain open.

\begin{figure}[hbp]
	\begin{center}
		\includegraphics[width=0.95\linewidth]{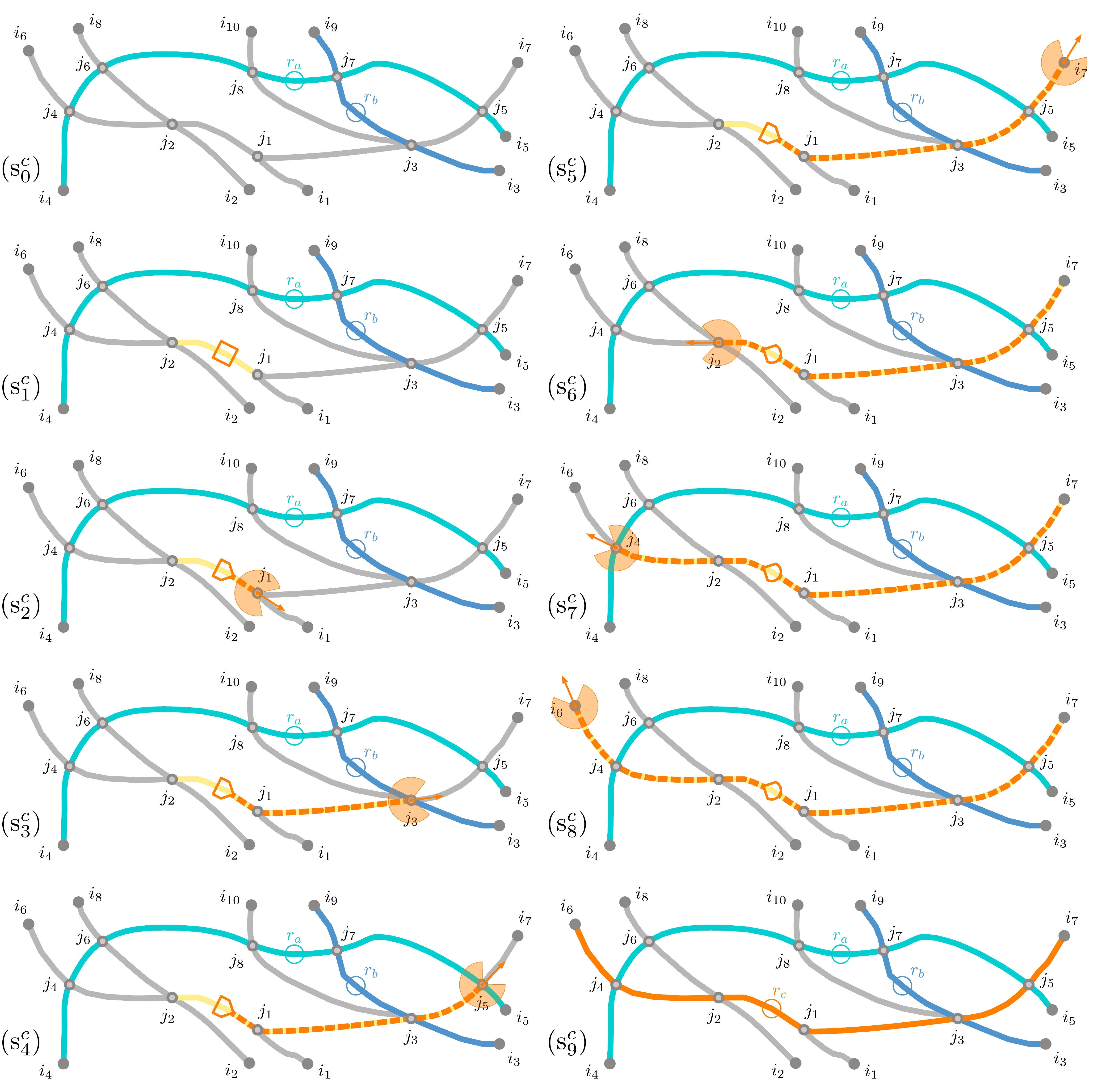}
	\end{center}
	\caption[State-of-the-art construction paradigm for configurations of streets]{%
			\label{OEMSSOUSN/fig/streetsegments/joinedsequence/notionalexamples}%
		State-of-the-art construction paradigm for configurations of streets:
		the frames ($\sotaStrStp[\ast]{\star}$) show
		this paradigm steps
		for constructing street $r_{c}$
		on street map ($\mathrm{m}$)
		from the notional example in Figure~\ref{OEMSSOUSN/fig/introduction/notionalexamples}.
		Our illustration assumes that
		streets $r_{a}$ and $r_{b}$ were constructed previously
		and that
		the remaining streets
		will be constructed afterward.
		The superscript and subscript of
		each frame label
		indicate
		the street under construction and the step order,
		respectively.
		Each street-segment in colour already belongs to a street:
		when the colour is vivid and the line is solid,
		the street was committed;
		when the colour is pallid or the line is vividly dashed,
		the street is under construction.
		Each street-segment in grey is a candidate for belonging to a new street.
		We attribute to each street a particular colour.
		The construction goes like this.
		\textsl{Initial stage} ($\sotaStrStp[c]{0}$):
			no street-segment is yet assigned to street $r_{c}$.
		\textsl{Seeding step} ($\sotaStrStp[c]{1}$):
			pick at random one candidate street-segment
			---
			the seed street-segment is in pallid orange and marked with an orange bold square.
		\textsl{Orientation step} ($\sotaStrStp[c]{2}$):
			orient at random the seed street-segment and move toward the head junction
			---
			the square mark is now a pentagonal \textit{``home plate''} indicating the orientation,
			the path moved along
			from the mark to the head junction $j_{1}$ is now vividly dashed,
			the excluding pie aligned
			with the incoming tangent
			at $j_{1}$
			immediately
			identifies along which outgoing street-segments the street might continue
			(see Figure~\ref{OEMSSOUSN/fig/joinprinciple/deflectionangle/notionalexamples}).
		\textsl{Appending loop steps} ($\sotaStrStp[c]{3}$)--($\sotaStrStp[c]{5}$):
			arbitrarily continue
			at the head junction
			along any valid outgoing street-segment
			(see Figure~\ref{OEMSSOUSN/fig/joinprinciple/deflectionangle/notionalexamples})
			while applicable
			---
			at $j_{1}$
			the street might continue toward either $i_{1}$ or $j_{3}$
			(see Figure~\ref{OEMSSOUSN/fig/joinprinciple/deflectionangle/notionalexamples}$\mathrm{a}_{2}$),
			the latter choice was arbitrarily taken;
			at $j_{3}$
			the street might continue toward either $i_{3}$ or $j_{5}$,
			as the former choice was no more possible
			only the latter could be taken;
			at $j_{5}$
			the street can only continue toward $i_{7}$;
			at $i_{7}$
			the street can no more continue
			so that the recursion ended.
		\textsl{Inverting step} ($\sotaStrStp[c]{6}$):
			move toward the tail junction and formally invert orientation
			---
			the forward recursion lets now place to a backward recursion,
			the pentagonal mark
			has flipped its orientation
			and
			has rounded its tail
			to mark the epoch.
		\textsl{Prepending loop steps} ($\sotaStrStp[c]{7}$)--($\sotaStrStp[c]{8}$):
			arbitrarily continue
			at the tail
			(formal head)
			junction
			along any valid outgoing street-segment
			(see Figure~\ref{OEMSSOUSN/fig/joinprinciple/deflectionangle/notionalexamples})
			while applicable
			---
			at $j_{2}$
			the street might continue toward either $j_{4}$ or $j_{6}$
			(see Figure~\ref{OEMSSOUSN/fig/joinprinciple/deflectionangle/notionalexamples}$\mathrm{b}_{2}$),
			the former choice was arbitrarily taken;
			at $j_{4}$
			the street can only continue toward $i_{6}$;
			at $i_{6}$
			the street can no more continue
			so that the
			backward
			recursion ended.
		\textsl{Commit step} ($\sotaStrStp[c]{9}$):
			commit the new achieved street
			and
			loop forward to build the next street until applicable
			---
			the now achieved street $r_{c}$ is in solid line,
			its mark is an unbold circle,
			and it has a label;
			retrospectively,
			this step leads to \textsl{Initial stage} ($\sotaStrStp[d]{0}$)
			for the next street $r_{d}$
			while
			the above \textsl{Initial stage} ($\sotaStrStp[c]{0}$)
			appears
			to result from \textsl{Commit step} ($\sotaStrStp[b]{6}$)
			for the previously committed street $r_{b}$;
			the construction of streets
			loops until no more street-segment is unassigned.
		The arbitrary choices in \textsl{Appending} and \textsl{Prepending loop steps}
		are actually join principles
		(see ``\nameref{sec/modelization/dynamics/streetperspective}'' section).
		Supplementary Animation~{A\ref*{OEMSSOUSN/ani/streetsegments/joinedsequence/notionalexamples}}
		(Additional file~\ref{OEMSSOUSN/ani/streetsegments/joinedsequence/notionalexamples})
		shows
		a complete construction
		of the configuration of streets
		on street map ($\mathrm{m}$).
		}
\end{figure}

Three paradigms based on deflection angles have been mainly used to generate information networks.
Basically these paradigms are nonoverlapping walks governed by a \emph{join principle}.
The \emph{every-best-fit join principle} \citep{PortaTNAUSDA2006,BJiangSZhaoJYin2008}
acts at every junction
by joining its street-segment pairs in increasing order of their deflection angles,
until applicable.
The \emph{self-best-fit join principle} \citep{MPViana2013,BJiangSZhaoJYin2008}
and \emph{self[-random]-fit join principle} \citep{BJiangSZhaoJYin2008}
act sequentially on growing streets,
until applicable,
by randomly seeding them with a not-yet-selected street-segment
before recursively appending,
until applicable,
one of the not-yet-appended street-segments.
The self join principles differ only in the choice of the not-yet-selected street-segment to append.
Figure~\ref{OEMSSOUSN/fig/streetsegments/joinedsequence/notionalexamples} illustrates
how the inner recursion can construct an entire street;
supplementary Animation~{A\ref*{OEMSSOUSN/ani/streetsegments/joinedsequence/notionalexamples}}
(Additional file~\ref{OEMSSOUSN/ani/streetsegments/joinedsequence/notionalexamples})
shows
how the full machinery can achieve a complete configuration of streets.
The self-best-fit join principle
selects the one forming the smallest deflection angle;
the self[-random]-fit join principle
selects at random.
By construction,
these three joint principles fall into two categories.
The every-best-fit join principle is local and almost deterministic%
\endnote{%
The every-best-fit join principle is
almost deterministic
in the sense that it resolves at random
the very rare occurrences of equality between deflection angles.}%
\label{OEMSSOUSN/en/almostdeterministic};
the two self join principles are global and random.
The latters clearly differ nevertheless in the degree of their randomness.
Unsurprisingly,
due to their walk-oriented construction,
the two self join principles have appeared,
against well-founded cadasters
and transportation traffic in terms of correlation,
more realistic \citep{BJiangSZhaoJYin2008}.
They thusly show that
the deflection angle is a suitable parameter
for generating information networks.
However,
the same walk-oriented construction renders them not easily tractable.
In short,
even though it provides a suitable parameter,
the state-of-the-art approach can not be used to build an easily tractable dynamics.

\subsubsection*{\addcontentsline{toc}{subsubsection}{A junction is a matching of street-segments}A junction is a matching of street-segments}

For information networks,
edges are junctions,
essentially
an exclusive set of singletons and pairs of street-segments
that are isolated or paired according to the ongoing streets.
By exclusive we mean that a street-segment can only belong either to one singleton or to one pair.
Such a set is,
in graph theory,
a \emph{matching} \citep{PEMMARJU}.
To the best of our knowledge,
this is the first work that mentions this perspective.

\begin{figure}[hbtp]
	\begin{center}
		\includegraphics[width=0.95\linewidth]{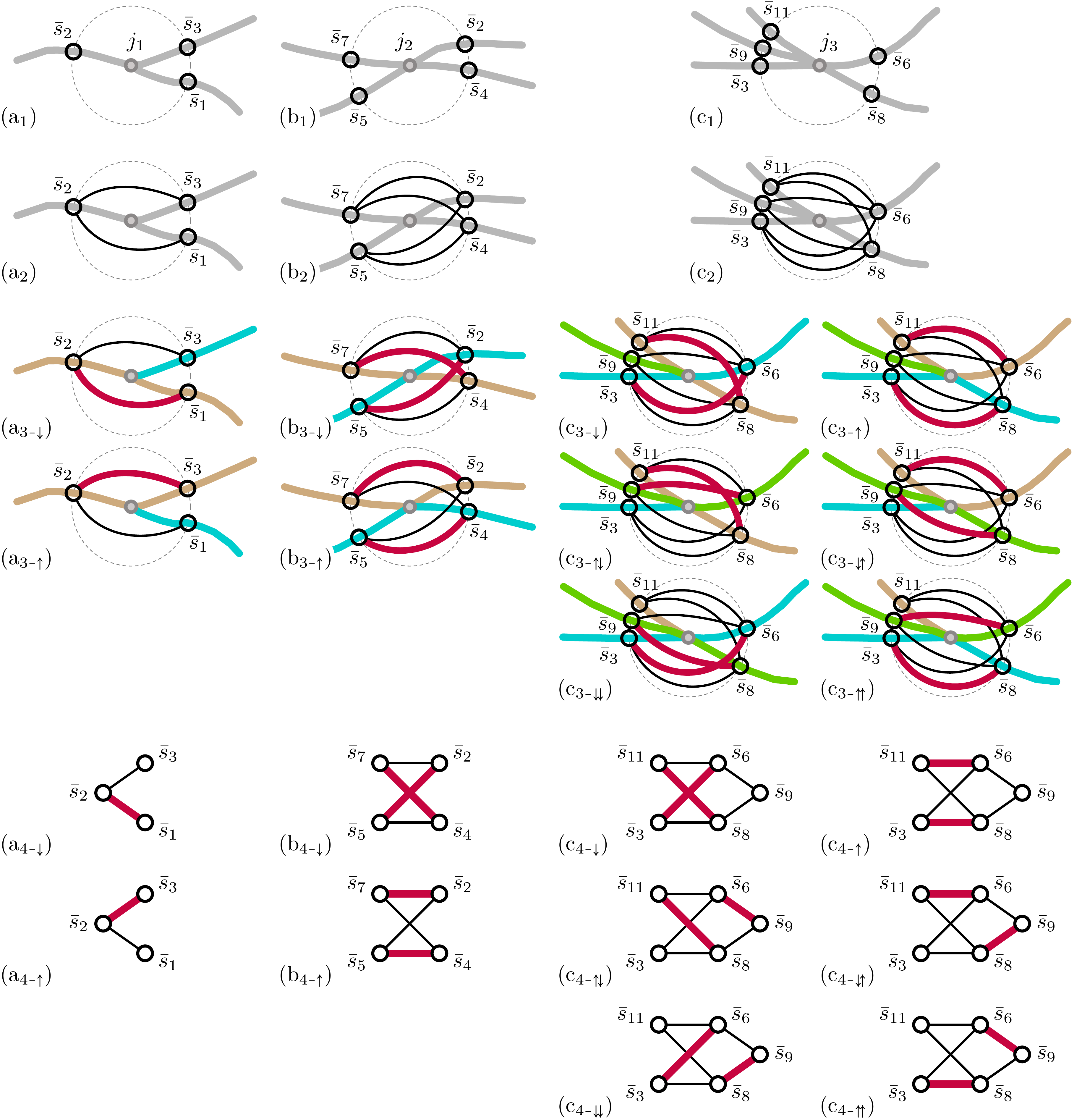}
	\end{center}
	\caption[Set of maximum matchings associable to a junction]{%
			\label{OEMSSOUSN/fig/junctions/matching/maximal/notionalexamples}%
		Set of maximum matchings associable to a junction:
		the columns ($\mathrm{a}_{\star}$), ($\mathrm{b}_{\star}$), and ($\mathrm{c}_{\star}$)
		outline
		why and how
		we can associate
		a set of maximum matchings
		to any junction
		through
		junctions $j_{1}$, $j_{2}$, and $j_{3}$,
		respectively,
		from the notional example in Figure~\ref{OEMSSOUSN/fig/introduction/notionalexamples}.
		The subfigures actually organize themselves in a table:
		each column corresponds to a
		notional
		junction;
		each row corresponds to
		a step of our outline.
		\textsl{First step}:
			associate to each street-segment a node
			---
			we can place
			each node
			at the intersection of the street-segment
			with a circle centred at the junction
			so that
			each graph is a circular graph.
		\textsl{Second step}:
			link each node to any node along which it might continue its way
			(see Figure~\ref{OEMSSOUSN/fig/joinprinciple/deflectionangle/notionalexamples})
			---
			each graph is actually
			a graph representation of the pairable street-segments.
		\textsl{Third step}:
			enumerate all the street layouts
			which could be achieved
			as part of a configuration of streets
			according to
			the state-of-the-art
			(see Figure~\ref{OEMSSOUSN/fig/streetsegments/joinedsequence/notionalexamples})
			---
			the construction of the street layouts
			follow
			the scheme
			used in Figure~\ref{OEMSSOUSN/fig/streetsegments/joinedsequence/notionalexamples},
			while
			the subgraphs link
			the so paired street-segment-nodes
			with a fat red edge.
		\textsl{Fourth step}:
			by representing the subgraphs in canonical form,
			we immediately realize that
			\textsl{Step three}
			actually
			enumerate all the subgraphs
			with the maximum number
			of non-adjacent edges,
			namely,
			all the maximum matchings
			---
			this completes our outline.
		Notice that
		the subfigure labels
		in rows 3 and 4
		enumerate the maximum matchings
		with balanced ternary numbers
		using
		down-spin ($\downarrow$),
		nil-spin ($0$),
		and
		up-spin ($\uparrow$)
		as ternary digits
		\citep{TAOCPSA}.
		This enumeration offers
		between
		maximum matchings of junctions joining three or four street-segments
		and
		Ising spin states
		(down $\downarrow$ and up $\uparrow$)
		a close analogy
		as
		junctions $j_{1}$ and $j_{2}$ exemplify well here.
		This analogy appears to hold as well for our less typical junction $j_{3}$.
		Figure~\ref{OEMSSOUSN/fig/junctions/flipflapflopsequence/notionalexamples}
		uses a simpler but more visual
		enumeration
		based on the regular polygons
		---
		extended with the degenerate regular digon.
		Each corner
		(side)
		represents then a maximum matching.
		}
\end{figure}

The graph theory perspective can apply on junctions as follows.
First,
inspired by the dual network representation of urban street networks,
we may represent every street-segment attached to a junction by a node.
Let us
put each node at the intersection of its associated street-segment
with a circle centred at the junction.
Second,
we may link pair of nodes whose associated street-segments
have a deflection angle smaller than the deflection angle threshold.
Figure~\ref{OEMSSOUSN/fig/junctions/matching/maximal/notionalexamples}
illustrates
in its two first rows
these two steps
for three realistic junctions.
The resulting graph clearly depends on the deflection angle threshold:
when it is set to the flat angle~$\pi$, the graph is a complete graph;
when it is set to the zero angle $0$, the graph is an empty graph;
otherwise
the graph is an incomplete graph.
We will call such a graph a \emph{junction graph}.
In general,
a junction graph has no direct application for our purpose in the sense that
any bunch of edges that share a common node
(or adjacent edges)
corresponds to a set of overlapping streets.
In practice,
we want a graph without any adjacent edge
so that the graph corresponds to a set of nonoverlapping streets.
Such a graph is,
for a given junction,
a \emph{matching subgraph}
(or matching for short)
\citep{PEMMARJU}
of its a junction graph.
In short,
we are interested by the set of matchings of the junction graphs.
The number of matchings of a graph is called the \emph{Hosoya index} \citep{HHosoya1971}.
We will denote the Hosoya index of the junction graph of a junction $j$ by $\mathscr{Z}_{j}$.
Also notice that
a matching can be saturated in the sense that
it cannot be expanded to another matching by adding any edge of the underlying graph.
Such a matching is called a \emph{maximal matching} \citep{PEMMARJU}.
Figure~\ref{OEMSSOUSN/fig/junctions/matching/maximal/notionalexamples}
gives
in its fourth row
the set of maximum matchings
we can derive
for each of its junctions.
We will denote the number of maximal matchings of the junction graph of a junction $j$ by $\mathscr{Y}_{j}$;
we have
\begin{equation}
	{1}\leqslant{\mathscr{Y}_{j}}\leqslant{\mathscr{Z}_{j}}
	.
\end{equation}

Let us now describe the previous generating paradigms in terms of matchings.
The \textit{``named street''} paradigm selects the matching as implicitly recorded in cadasters.
The every-best-fit join principle chooses
for each junction
the maximal matching which is optimal in terms of deflection angle distribution.
The two self join principles operate at every junction on the set of matchings by successive visits.
This becomes more apparent when we interpret their concrete implementations as
nonoverlapping walks that haphazardly visit every junction several times.
Each visit either steps forward or terminates the walk,
that is,
each visit selects a subset of matchings.
This selection process reveals itself
in
supplementary Animation~{A\ref*{OEMSSOUSN/ani/streetsegments/joinedsequence/notionalexamples}}
(Additional file~\ref{OEMSSOUSN/ani/streetsegments/joinedsequence/notionalexamples}).
For the self-best-fit join principle,
the move is optimal in terms of deflection angle;
for the self[-random]-fit join principle,
the move is random.
Over the visits
the subset of matchings decreases until it contains only one matching.
This remaining matching is a maximal matching
since every walk terminates only when no more street-segment is attachable.
Figure~\ref{OEMSSOUSN/fig/junctions/matching/maximal/notionalexamples}
draws
in its third row
the end results of these repeated visits
along their maximum matching
for each of its junctions.
Actually,
Figure~\ref{OEMSSOUSN/fig/junctions/matching/maximal/notionalexamples}
sketches,
through three realistic junctions,
why and how
to any junction corresponds a set of maximum matching.
The so isolated maximal matchings give the generated information network.
In other words,
the self join paradigms interpret themselves now
as an intricate haphazard fashion to pick for every junction a maximal matching.
Thusly,
the matching viewpoint allows us to slightly disentangle
the two most pertinent join paradigms.

\subsubsection*{\addcontentsline{toc}{subsubsection}{The single-junction-switch and -flip dynamics}The single-junction-switch and -flip dynamics}

The previous slightly untangled description
actually
leads to a disembodied form of the self join paradigms
with all their underlying principles removed.
This is exactly what an ergodic dynamics is about.
To the best of our knowledge,
no ergodic dynamics has been reported
so far
for generating configurations of streets.

To begin with,
let us deliberately ignore for a while the nonoverlapping walk machineries.
The self join paradigms reduce then
to choose for every junction a maximal matching regardless of the matchings of the other junctions.
So the elementary disembodied dynamics that occurs at junctions is
to set up
in an independent way
a maximal matching.
The new set up will generally change the maximal matching into another maximal one.
For clarity,
this dynamics does not alter
the urban street network
but rather transforms the information network
into another,
since the new maximal matching sets a new layout for some of the streets that cross the junction.
By now we are able to tell that this dynamics is ergodic.
An ergodic dynamics is a dynamics which from any state can reach any other state after a finite number of iterations.
It is indeed obvious that
we can get from any information network to any other
by changing one by one each of the maximal matchings by which the two information networks differ.
We coined this dynamics,
following the literature \citep{MCMSP}
and
as an obvious analogy to railroad switches,
the \emph{single-junction-switch} dynamics.
The restriction to consider only maximal matchings is inherited from the join principles.
This restriction is arbitrary in the sense that it is not actually imposed by physical constraints.
In fact,
the reasoning held above for the single-junction-switch dynamics
evidently
holds for any arbitrary choice of subset of matchings.
For completeness,
we coined the dynamics that involves all matchings
the \emph{single-junction-flip} dynamics.
Let us recap along these lines.
Assuming an entire urban street network,
the single-junction-switch dynamics is an ergodic dynamics which
switches
the maximal matching of a single junction into another maximal one,
while
the single-junction-flip dynamics is an ergodic dynamics which
flips
the matching of a single junction into another.

Using
the single-junction-switch or -flip dynamics
ensures that
our Metropolis algorithm fulfils the condition of ergodicity.
It remains however to specify how we select
from a given information network a new one which differs by only one dynamics step.
For the sake of simplicity,
we will only consider
the single-junction-switch dynamics
in the following.
The choice of the Metropolis algorithm imposes
\citep{MCMSP,DPLandauKBinderGMCSSP}
that
the selection probabilities $g(\mu\to\nu)$
for each possible new information network~$\nu$ after one dynamics step
are all chosen equal
---
the selection probabilities for all other information networks are set to zero.
For an entire urban street network,
one dynamics step enables to reach each of the $\mathscr{Y}_{j}-1$ new maximal matchings of every junction $j$.
Hence the number of possible new information networks that
we can reach after one dynamics step from a given information network
is the total number of maximal matchings
\begin{equation}
	\mathscr{N} = \sum_{j}\mathscr{Y}_{j}
\end{equation}
minus the number of junction $N$.
Therefore we count $\mathscr{N}-N$ non-zero selection probabilities $g(\mu\to\nu)$,
and each of them takes the value
\begin{equation}
	g(\mu\to\nu) = \frac{1}{\mathscr{N}-N}
	.
\end{equation}
In practice we can realize this selection in two easy steps.
First we pick at random a junction $j$ with probability proportional to ${\mathscr{Y}_{j}-1}$.
Then we choose at random a new maximal matching
among the ${\mathscr{Y}_{j}-1}$ possible new maximal matchings of junction $j$.

\subsection*{\addcontentsline{toc}{subsection}{Another variant of the single-spin-flip Metropolis algorithm ?}\label{sec/modelization/dynamics/variant}Another variant of the single-spin-flip Metropolis algorithm ?}

When all junctions have two maximal matchings,
the single-junction-switch dynamics is formally equivalent to
the single-spin-flip dynamics on the original Ising model.
Our two above achievements actually combine to give
another variation on the single-spin-flip Metropolis algorithm theme.
This algorithm is a computational interpretation of the Ising model.
A brief comparison provides
basic physical insights
and
a simple clue for a crossover as scaling varies.

\begin{figure}[hbp]
	\begin{center}
		\includegraphics[width=0.95\linewidth]{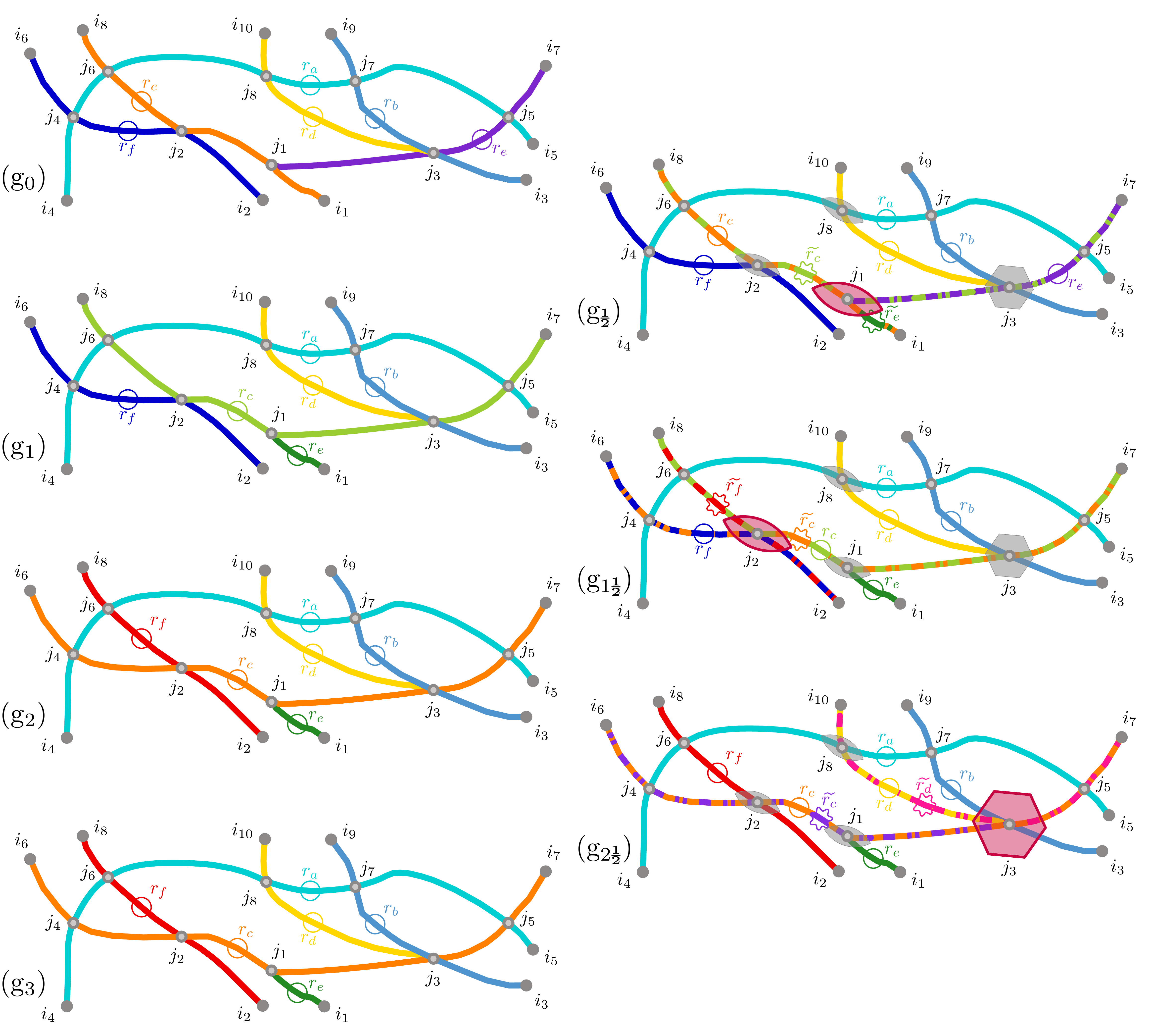}
	\end{center}
	\caption[Single-junction-switch Metropolis algorithm for configurations of streets]{%
			\label{OEMSSOUSN/fig/junctions/flipflapflopsequence/notionalexamples}%
		Single-junction-switch Metropolis algorithm for configurations of streets:
		the frames ($\mathrm{g}_{\star}$) show
		how this algorithm may evolve
		from the configuration of streets
		on street map ($\mathrm{m}^{\prime}$) to the one on street map ($\mathrm{m}$)
		from the notional example in Figure~\ref{OEMSSOUSN/fig/introduction/notionalexamples}.
		The subscript of each frame label
		indicate the generation time.
		The figure reads from top to bottom.
		It actually separates
		generated configurations of streets
		from intermediate computational steps
		in two complementary ways:
		it shifts
		(resp. epochs)
		the formers to the left column
		(resp. with integers)
		and
		the latters to the right column
		(resp. with half-integers).
		The right frames display more information.
		In particular
		each switchable junction
		is marked with an extended regular polygon
		whose the corners (sides) enumerate its maximum matchings
		(see Figure~\ref{OEMSSOUSN/fig/junctions/matching/maximal/notionalexamples}):
		when the polygon is in grey,
		the junction is resting;
		when the polygon is in red and notably bigger,
		the junction is switching.
		As for the state-of-the-art construction paradigm
		in Figure~\ref{OEMSSOUSN/fig/streetsegments/joinedsequence/notionalexamples},
		we attribute to each street a particular colour.
		In clear contrast,
		however,
		all streets are here fully constructed.
		The streets in dashed lines are not under construction but instead under challenge
		as follows.
		The dashing actually indicates
		coexistence of old streets with new ones.
		Each old street keeps its colour, its circle mark, and its label.
		Each new street emerges with a new color, a wavy-circle mark, and a tilded label.
		The new streets result from the new street layout at the switching junction.
		This change leads so to a new configuration of streets $\nu$ that competes with the old one $\mu$.
		This is the
		actual
		ongoing challenge.
		The Metropolis algorithm resolves such challenges by either accepting or rejecting
		change $\mu\to\nu$ with
		an acceptance
		ratio
		$A(\mu\to\nu)$
		in an optimal way.
		After a sufficient number of generations,
		the configurations of streets reach a prescribed statistical equilibrium
		---
		provided the equilibrium is sustainable.
		Our prescribed statistical equilibrium
		follows from the assumption that
		self-organized urban street networks are statistically self-similar.
		It is a Boltzmann-like distribution
		with a total amount of surprisal
		(information)
		instead of energy
		and
		the scaling as equilibrium parameter
		(see formula~\eqref{OEMSSOUSN/eq/SOUSN/surprisal/probability/formula}).
		Our odds of accepting or rejecting new configurations of streets
		favour the less surprising ones
		(see formulae~\eqref{OEMSSOUSN/eq/SOUSN/Metropolis/acceptanceratio}
			and~\eqref{OEMSSOUSN/eq/SOUSN/workinghypotesis/Metropolis/acceptanceratio}).
		The algorithm goes like this
		(see ``\nameref{sec/modelization/dynamics/variant/informalimplementation}'' section).
		Note first that
		only the four junctions $j_{1}$, $j_{2}$, $j_{3}$, and $j_{8}$ are actually switchable:
		junction
		$j_{3}$ has ${\mathscr{Y}_{3}=6}$ maximum matchings
		(see Figure~{\ref{OEMSSOUSN/fig/junctions/matching/maximal/notionalexamples}$\mathrm{c}$});
		junctions
		$j_{1}$, $j_{2}$, and $j_{8}$ have ${\mathscr{Y}_{1}=\mathscr{Y}_{2}=\mathscr{Y}_{8}=2}$ maximum matchings
		(see Figures~{\ref{OEMSSOUSN/fig/junctions/matching/maximal/notionalexamples}$\mathrm{a}$}
			and~{\ref{OEMSSOUSN/fig/junctions/matching/maximal/notionalexamples}$\mathrm{b}$});
		the remaining junctions $j_{\circ}$ have obviously ${\mathscr{Y}_{\circ}=1}$ maximum matching.
		So each move will first
		choose randomly one junction among junctions $j_{1}$, $j_{2}$, $j_{3}$, and $j_{8}$
		with probabilities $\tfrac{1}{8}$, $\tfrac{1}{8}$, $\tfrac{5}{8}$, and $\tfrac{1}{8}$,
		respectively.
		Each move will second pick uniformly at random a new maximum matching.
		There will be ${\mathscr{Y}_{3}-1=5}$ choices for $j_{3}$,
		and ${\mathscr{Y}_{1}-1}={\mathscr{Y}_{2}-1}={\mathscr{Y}_{8}-1}=1$ choice for $j_{1}$, $j_{2}$, and $j_{8}$.
		Each move will third calculate its change in amount of surprisal
		$\Delta{S}$
		in view to compute its acceptance ratio $A$.
		Here
		the surprisal changes
		at mid-steps
		($\mathrm{g}_{\oneonehalf}$),
		($\mathrm{g}_{1\oneonehalf}$),
		and
		($\mathrm{g}_{2\oneonehalf}$)
		are
		respectively
		$
		\Delta{\widetilde{S}}
			=
				\ln\frac{5}{9}
			$,
		$%
		\Delta{\widetilde{S}}
			=
				0
			$,
		and
		$%
		\Delta{\widetilde{S}}
			=
				\ln\frac{6}{5}
			$
		(see ``\nameref{sec/metropolis/workingassumptions}'' section).
		Ultimately each move will either accept or reject the change with probability $A$.
		The layout changes at ($\mathrm{g}_{\oneonehalf}$) and ($\mathrm{g}_{1\oneonehalf}$) are certain
		since they are less or equally surprising,
		the one at ($\mathrm{g}_{2\oneonehalf}$) is accepted with probability $\exp(-\widetilde{\lambda}\ln\frac{6}{5})$
		where $\widetilde{\lambda}$ is our effective equilibrium parameter
		(see formula~\eqref{OEMSSOUSN/eq/SOUSN/workinghypotesis/Metropolis/acceptanceratio}).
		Our illustration actually rejects the last move.
		Supplementary Animation~{A\ref*{OEMSSOUSN/ani/junctions/flipflapflopsequence/notionalexamples}}
		(Additional file~\ref{OEMSSOUSN/ani/junctions/flipflapflopsequence/notionalexamples})
		shows
		a longer sequence.
		}
\end{figure}

\subsubsection*{\addcontentsline{toc}{subsubsection}{Informal implementation}\label{sec/modelization/dynamics/variant/informalimplementation}Informal implementation}

As summary of our above results,
let us informally implement our adaptation of the Metropolis algorithm
to urban street networks as follows.
\begin{quote}
First,
we
choose randomly a junction $j$
with probability proportional to its number of maximal matchings $\mathscr{Y}_{j}$ minus $1$,
${\mathscr{Y}_{j}-1}$;
its street-segments will be laid out according to some maximal matching $\mathsf{M}_{j}$.
Second,
we
pick at random a new maximal matching $\widetilde{\mathsf{M}}_{j}$ not identical to $\mathsf{M}_{j}$
among the remaining ${\mathscr{Y}_{j}-1}$ available possibilities.
Third,
we
calculate the change in the total amount of surprisal $\Delta{S}$
that would result if we were to lay out this change to this junction.
Ultimately,
with acceptance probability
\begin{equation*}
	A=
		\begin{cases}
			\mathrm{e}^{-\lambda {\Delta{S}}} & \text{if } {\Delta{S}}>0\\
			1 & \text{otherwise},
		\end{cases}
\end{equation*}
either accept or reject the change.
\end{quote}
Properly speaking,
this informal implementation encodes
the single-junction-switch Metropolis algorithm
for urban street networks.
We let the reader to elaborate
the corresponding single-junction-flip Metropolis algorithm.
Meanwhile,
the reader may refer to
Figure~\ref{OEMSSOUSN/fig/junctions/flipflapflopsequence/notionalexamples}
and supplementary Animation~{A\ref*{OEMSSOUSN/ani/junctions/flipflapflopsequence/notionalexamples}}
(Additional file~\ref{OEMSSOUSN/ani/junctions/flipflapflopsequence/notionalexamples})
for illustration.

\subsubsection*{\addcontentsline{toc}{subsubsection}{A brief comparison with Ising models}\label{sec/modelization/dynamics/variant/comparison}A brief comparison with Ising models}

The single-junction-switch
(resp. single-junction-flip)
Metropolis algorithm
for our urban street network model
mimics
the classical single-spin-flip Metropolis algorithm
for Ising models \citep{MCMSP,DPLandauKBinderGMCSSP,JBerlinskyABHarrisGTPSM,DJCMacKayITILA}.
Nonetheless our model differs from them
in three basic aspects:
\begin{enumerate}[i), ref=(\roman*)]
	\item Our model is driven by scaling and surprisal (information)
		whereas Ising models are driven by temperature and energy.
		The parallel scaling-information versus temperature-energy
		$(\lambda,{S})\leftrightarrow(\beta,{E})$
		pours into the discipline
		the all maturity of thermodynamics and statistical physics.
		This parallel is actually superseded by
		the maximum entropy formalism
		\textit{``in a disembodied form with all the physics removed''} \citep{ETJaynesPTLS}.
		This formalism provides,
		for instance,
		numerical tools to compute
		for any information network measure \citep{MEJNewmanNI,PortaTNAUSDA2006}
		its linear response to arbitrary small scaling changes,
		namely
		its specific-heat-capacity-like coefficient
		or
		\emph{susceptibility}
		\citep{MCMSP,WTGrandyJrFSMET,ETJaynesPTLS}.
	\item Junctions are nodes of a finite arbitrary planar graph
		while spins are classically attached to sites of an \textit{``infinite''} regular lattice.
		Finiteness means that collective phenomena will get smoother.
		Arbitrariness renders our model closer to the Ising spin-glass models for which
		the values of the spin-spin interactions are no more constant
		but random \citep{MCMSP,DPLandauKBinderGMCSSP,DJCMacKayITILA}.
		Collective phenomena in Ising spin-glass models are
		more subtle and more intricate \citep{MCMSP,DPLandauKBinderGMCSSP}.
	\item The distribution of maximal matchings
		(resp. matchings)
		among junctions is heterogeneous
		while the distribution of spins among sites is classically homogeneous.
		That is,
		junctions have different numbers of maximal matchings
		(resp. numbers of matchings (Hosoya indices))
		while spins have classically the same number of states or the same dimension.
		Because the number of street-segments attached to a junction is mostly three or four,
		the distribution of matching is expected to be statistically homogeneous with a bell-like distribution.
		This contributes to make our model even closer to the Ising spin-glass models.
\end{enumerate}
On the other hand,
it is noteworthy that
the Ising models have became a toy model to crack
\emph{phase transition} and \emph{crossover} phenomena \citep{JBerlinskyABHarrisGTPSM,DJCMacKayITILA}.
This raises the obvious question whether
our model may actually undergo a crossover
as scaling varies:
\begin{enumerate}[resume*]
	\item\label{pt/modelization/dynamics/variant/comparison/crossover}
	Our model experiences,
	as scaling increases,
	an ultra-small- to small-world crossover
	around the scaling value of $3$.
	The small-world effect is in effect a statement
	on geodesic
	(or shortest)
	distances between node pairs \citep{MEJNewmanNI}.
	Their mean $\ell$ behaves
	in small-world networks
	as the logarithm of the number of nodes $N$,
	${\ell\sim\ln{N}}$
	\citep{MEJNewmanNI}.
	The small-world effect becomes extreme
	in scale-free networks
	as the scaling $\lambda$ get smaller than $3$
	\citep{RCohenSHavlin2003}:
	the \emph{mean geodesic distance} $\ell$ behaves as ${\ell\sim\ln\ln{N}}$ when ${2<\lambda<3}$,
	as ${\ell\sim\ln{N}/\ln\ln{N}}$ at ${\lambda=3}$,
	and
	as ${\ell\sim\ln{N}}$ for ${3<\lambda}$.
	Thusly,
	since our generated information networks are scale-free,
	our model effectively undergoes a small-world crossover as scaling varies.
	Clearly
	its manifestation relies on the behaviour of the number of nodes $N$.
	A more substantial analytical work is however beyond the scope of the present paper.
	Meanwhile,
	notice that
	a geodesic distance counts
	in our context
	how many changes of street are required for a particular journey.
	That is,
	the mean geodesic distance reflects how rapidly on average city-dwellers can travel.
	Accordingly,
	the crossover diagram of the mean geodesic distance
	interprets itself as an efficiency diagram.
	This means,
	for instance,
	that our approach provides
	a method to analyze the relative efficiency of an actual configuration of streets.
\end{enumerate}

\section*{\addcontentsline{toc}{section}{Equilibrium Metropolis simulations}\label{sec/metropolis}Equilibrium Metropolis simulations}

What we have achieved in the previous section is
adapting to
unplanned or
self-organized urban street networks
the Metropolis algorithm.
Now,
in this section,
we exercise this adaptation
in a case study.
As case study,
we select
the urban street network of Central London (United Kingdom),
which is a classical example of self-organized urban street network \citep{GreatStreets1993}.

\subsection*{\addcontentsline{toc}{subsection}{Working assumptions}\label{sec/metropolis/workingassumptions}Working assumptions}

For the sake of illustration,
we have made two suppositions.
First,
we have assumed that streets predominate junctions.
Second,
we have described the mesoscopic streets as asymptotic agent systems driven
by social interactions \citep{YDover2004,JBenoitOPSOUSN2019,SESOPLUSN}.
According to this agent model,
a \emph{number of vital connections} $\upsilon$ dominates among
the possible numbers of connection between agents.
So that,
the number of configurations $\Omega_{s_{\mu}}$ of street $s_{\mu}$
in
configuration of streets
$\mu$
becomes
proportional to a power of its number of junctions $n_{s_{\mu}}$ \citep{JBenoitOPSOUSN2019,SESOPLUSN};
we have
\begin{equation}
	\Omega_{s_{\mu}}\propto{n_{s_{\mu}}^{2\upsilon}}
	.
\end{equation}
So
the total amount of surprisal $S_{\mu}$ \eqref{OEMSSOUSN/eq/SOUSN/surprisal/probability/def/amount/total}
in
configuration of streets
$\mu$
becomes
\begin{equation}\label{OEMSSOUSN/eq/SOUSN/workinghypotesis/surprisal/amount/total}
	S_{\mu}=2\upsilon\sum_{s_{\mu}}\ln{n_{s_{\mu}}}
\end{equation}
up to an irrelevant constant;
the sum is over the streets $s_{\mu}$ of
configuration of streets~$\mu$.
Thusly our working assumptions bring out
an \emph{effective scaling exponent}~$\widetilde{\lambda}$
along an \emph{effective total amount of surprisal} $\widetilde{S}_{\mu}$;
we read
\begin{equation}\label{OEMSSOUSN/eq/SOUSN/workinghypotesis/rescaling/def}
	\widetilde{\lambda} = 2\lambda\upsilon
	\qquad\text{and}\qquad
	\widetilde{S}_{\mu} = \sum_{s_{\mu}}\ln{n_{s_{\mu}}}
	.
\end{equation}
The
corresponding
effective Metropolis acceptance ratio is literally
the tilde version of
formula
\eqref{OEMSSOUSN/eq/SOUSN/Metropolis/acceptanceratio};
we get
\begin{equation}\label{OEMSSOUSN/eq/SOUSN/workinghypotesis/Metropolis/acceptanceratio}
	A(\mu\to\nu)=
		\begin{cases}
			\mathrm{e}^{-\widetilde{\lambda} {(\widetilde{S}_{\nu}-\widetilde{S}_{\mu})}} &
						\text{if } {\widetilde{S}_{\nu}-\widetilde{S}_{\mu}}>0\\
			1 & \text{otherwise}.
		\end{cases}
\end{equation}
Figure~\ref{OEMSSOUSN/fig/junctions/flipflapflopsequence/notionalexamples}
along with
supplementary Animation~{A\ref*{OEMSSOUSN/ani/junctions/flipflapflopsequence/notionalexamples}}
(Additional file~\ref{OEMSSOUSN/ani/junctions/flipflapflopsequence/notionalexamples})
show
how our Metropolis adaptation
can actually generate
a sequence of configurations of streets.

\begin{figure}[hbp]
	\begin{center}
		\includegraphics[width=0.95\linewidth]{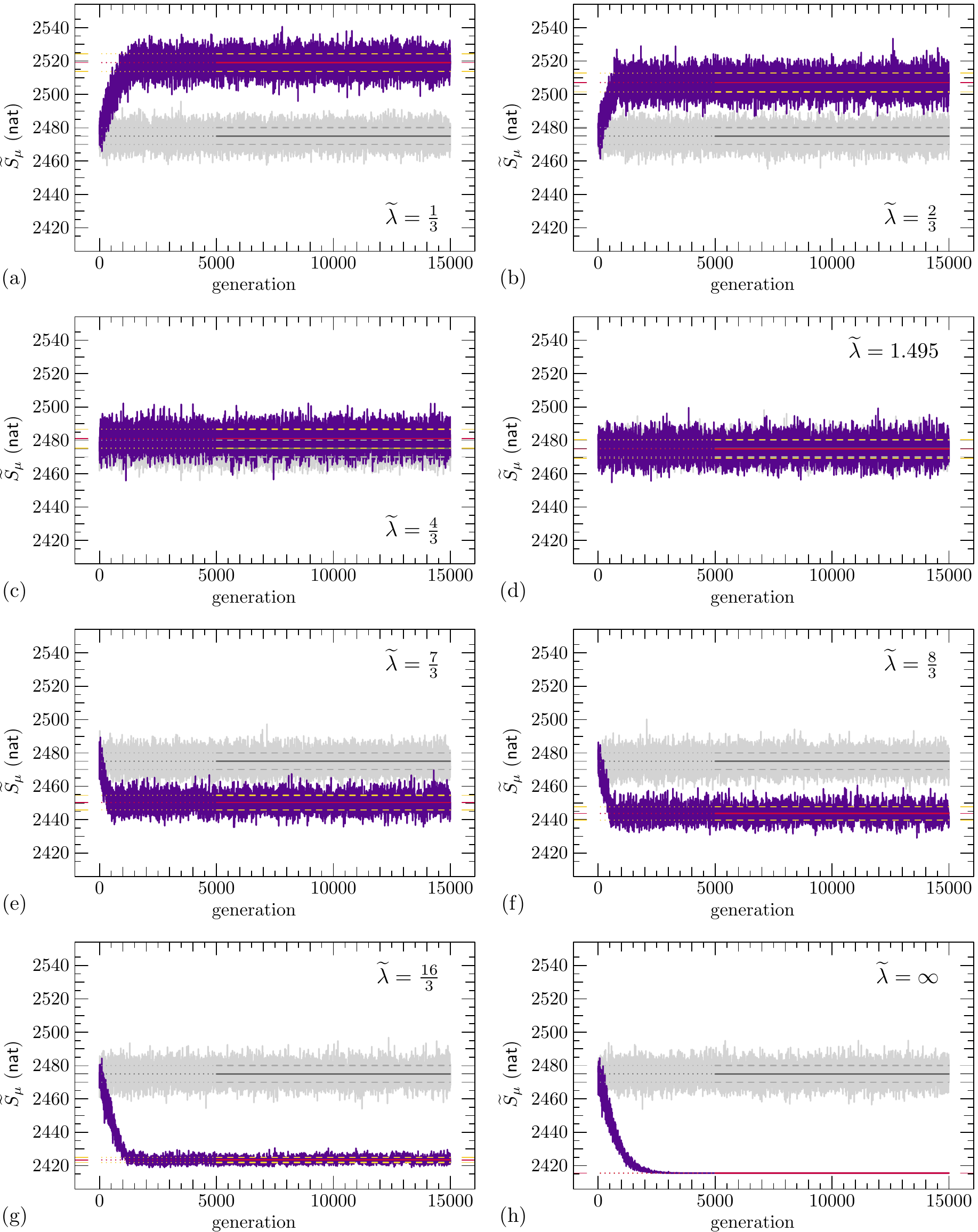}
	\end{center}
	\caption[Typical single-junction-switch Metropolis generation series for Central London]{%
			\label{OEMSSOUSN/fig/surprisal/Metropolis/equilibria/CentralLondon/series}%
		Typical single-junction-switch Metropolis
		generation series
		for Central London
		(United Kingdom):
		the foreground purple generation series plot,
		for different effective scaling exponents $\widetilde{\lambda}=2\lambda\upsilon$,
		typical simulations starting from a self-fit output;
		the background light-grey generation series plot,
		according to the same \textit{modus operandi},
		typical sequences of self-fit outputs.
		Each horizontal solid line drawn along a generation series plot represents its in-equilibrium mean value,
		the accompanying horizontal dashed lines indicate the associated standard-deviation bounds.
		Coloured lines, Metropolis generation series; greyed lines, sequences of self-fit outputs.
		The equilibria are assumed reached after the $5\,000^{\:\mathrm{th}}$ generation.
		The following {annealing parameters} were used to algebraically cool down to the desired $\widetilde{\lambda}$:
		$\widetilde{\lambda}_{0}=1.495$, $\epsilon={10^{-3}}$ and $m=500$.
		(%
		The starting cooling value
		$\widetilde{\lambda}_{0}$
		was chosen
		by hand
		so that the associated Metropolis equilibrium
		approaches on average self-fit outputs
		---
		as illustrated in (d);
		the special case $\widetilde{\lambda}=\infty$
		in (h)
		corresponds to a full \textit{`simulated annealing'} process \citep{WHPress2007,GSL}.%
		)%
		}
\end{figure}

\subsection*{\addcontentsline{toc}{subsection}{Equilibria}\label{sec/metropolis/equilibria}Equilibria}

\subsubsection*{\addcontentsline{toc}{subsubsection}{Single-junction-switch Metropolis generation series vs. self-fit outputs}\label{sec/metropolis/equilibria/switching}Single-junction-switch Metropolis generation series vs. self-fit outputs}

Central London
offers,
as shown in Figure~\ref{OEMSSOUSN/fig/surprisal/Metropolis/equilibria/CentralLondon/series},
single-junction-switch Metropolis generation series that come to equilibria.
The equilibria were attained
from self-fit outputs
through a basic algebraic \emph{annealing schedule} \citep{MCMSP,WHPress2007,GSL}.
To paraphrase:
increase
(resp. decrease)
the control effective scaling exponent $\widetilde{\lambda}_{c}$ to $\widetilde{\lambda}_{c}(1+\epsilon)$
(resp. $\widetilde{\lambda}_{c}/(1+\epsilon)$)
after every $m$ accepted/rejected single-junction-switch
moves
up
(resp. down)
to the desired equilibrium effective scaling exponent $\widetilde{\lambda}$;
the initial control effective scaling exponent $\widetilde{\lambda}_{0}$
and the parameters $\epsilon$ and $m$ are determined by experiment.
This annealing schedule allowed us
to reach equilibria
for a range of effective scaling exponent $\widetilde{\lambda}$ values
large enough
to capture the features of our system as follows.

\begin{figure}[hbp]
	\begin{center}
		\includegraphics[width=0.95\linewidth]{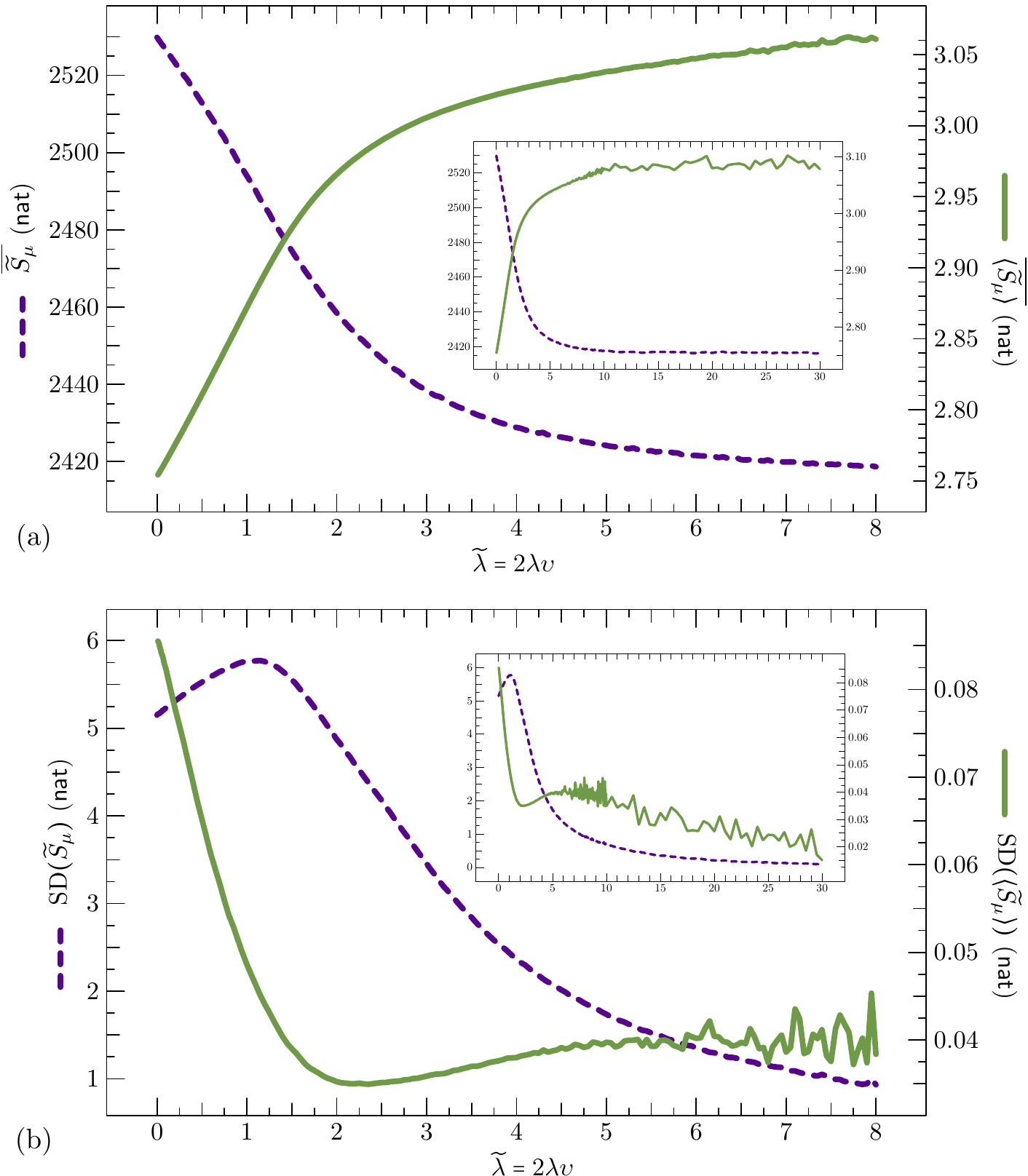}
	\end{center}
	\caption{\label{OEMSSOUSN/fig/surprisal/Metropolis/equilibria/CentralLondon/behaviour}%
		Effective
		total and average amounts of surprisal
		versus effective scaling exponent
		for
		Central London (United Kingdom):
		green solid lines plot the mean (top) and standard-deviation (bottom) of
		the effective average amount of surprisal $\langle\widetilde{S}_{\mu}\rangle$;
		purple dashed lines plot the mean (top) and standard-deviation (bottom) of
		the effective total amount of surprisal $\widetilde{S}_{\mu}$;
		the insets show their asymptotic behaviours.
		The annealing parameters are the same as
		in Figure~\ref{OEMSSOUSN/fig/surprisal/Metropolis/equilibria/CentralLondon/series};
		the equilibria were assumed reached after the $5\,000^{\:\mathrm{th}}$ generation
		as in Figure~\ref{OEMSSOUSN/fig/surprisal/Metropolis/equilibria/CentralLondon/series};
		the mean and standard-deviation values were computed over $250\,000$
		in-equilibrium
		generations
		and averaged over $10$ simulations.
		(%
		We attribute
		the noise that waves the asymptotic branches
		to
		the poor quality of our map data.\endnotemark[\ref{OEMSSOUSN/en/noise}]%
		)%
		}
\end{figure}

The
single-junction-switch Metropolis
generation series
exhibited
in Figure~\ref{OEMSSOUSN/fig/surprisal/Metropolis/equilibria/CentralLondon/series}
show
at least
four noticeable properties:
\begin{enumerate}[i)]
	\item The sustained Metropolis equilibria (a-f)
		are clearly comparable to the self-fit outputs
		in terms of order of magnitude
		of their means and fluctuations.
		This property holds,
		as shown Figure~\ref{OEMSSOUSN/fig/surprisal/Metropolis/equilibria/CentralLondon/behaviour},
		within a window grossly comprised between $1$ and $5$.
		We must always bear in mind that
		scaling exponents of real-world networks are typically comprised between $2$ and $3$ \citep{MEJNewmanNI}.
	\item The ground state,
		namely the sustained Metropolis equilibrium~(h) attained for ${\widetilde{\lambda}=\infty}$,
		lays below the self-fit outputs by about eleven times their standard-deviation.
		The ground state was obtained
		through a \textit{`simulated annealing'} \citep{WHPress2007,GSL}.
	\item The sustained Metropolis equilibrium~(a) shows that there also exist equilibria
		that detach significantly from the self-fit outputs from above.
	\item The sustained Metropolis equilibrium~(d) shows
		that the single-junction-switch Metropolis algorithm can mimic quite well
		sequences of self-fit outputs.
\end{enumerate}
These four properties lead us to claim that
the single-junction-switch Metropolis algorithm generates series that
are consistent
with the self-fit outputs.

The effective total and average amounts of surprisal,
$\widetilde{S}_{\mu}$ and $\langle\widetilde{S}_{\mu}\rangle$ respectively,
exhibit
in Figure~\ref{OEMSSOUSN/fig/surprisal/Metropolis/equilibria/CentralLondon/behaviour}
at least two promising properties:
\begin{enumerate}[i)]
	\item Their means and their standard-deviations vary smoothly
		as a function of the effective scaling exponent $\widetilde{\lambda}$
		at least for values smaller than $4.5$.
		For greater effective scaling exponent $\widetilde{\lambda}$ values,
		our simulations get subject to noise:
		the effective total amounts of surprisal $\widetilde{S}_{\mu}$
		and its standard-deviation $\stSD(\widetilde{S}_{\mu})$
		continue to vary smoothly while they tend asymptotically to a constant;
		however,
		their average counterparts $\langle\widetilde{S}_{\mu}\rangle$ and $\stSD(\langle\widetilde{S}_{\mu}\rangle)$
		experience noisy variations.%
\endnote{%
We attribute the noise to the poor quality of our map data
in their small streets
and
in their simplification of the junctions.
The variations of $\widetilde{S}_{\mu}$ and $\stSD(\widetilde{S}_{\mu})$ remains relatively smooth because
the Metropolis acceptance ratio \eqref{OEMSSOUSN/eq/SOUSN/workinghypotesis/Metropolis/acceptanceratio}
tends to smooth $\widetilde{S}_{\mu}$ itself by rejecting the inappropriate states
--- among them there are the inappropriate states resulting from \textit{``corrupted''} data.
By contrast,
the variation of $\langle\widetilde{S}_{\mu}\rangle$ and $\stSD(\langle\widetilde{S}_{\mu}\rangle)$
are not smoothed by the Metropolis algorithm in any manner.
Furthermore,
\textit{``corrupted''} layout at junctions cannot be rejected
because
(i)
our working assumptions do not take into account junctions in the computation of $\widetilde{S}_{\mu}$
and
because
(ii)
the single-junction-switch dynamics cannot break them since it only allows maximal layouts
---
this becomes evident as soon as the single-junction-flip dynamics is used
since then $\langle\widetilde{S}_{\mu}\rangle$ and $\stSD(\langle\widetilde{S}_{\mu}\rangle)$
vary almost smoothly along their respective asymptotic branch.%
}\phantomsection\label{OEMSSOUSN/en/noise} 
	\item They all experience a noticeable change of behaviour
		within the window comprised between $1$ and $4$.
		The means of $\widetilde{S}_{\mu}$ and $\langle\widetilde{S}_{\mu}\rangle$
		experience both a change of rate that leads them to their respective asymptotic plateau.
		The standard-deviations $\stSD(\widetilde{S}_{\mu})$ reach a maximum at right of~$1$
		before decreasing towards an asymptotic plateau;
		the standard-deviations $\stSD(\langle\widetilde{S}_{\mu}\rangle)$
		has the left profile of a Mexican-hat
		--- left shape of a biquadratic curve ---
		with a minimum around $2$.
\end{enumerate}
The latter property strongly suggests that
the relevant physics of our system occurs
within the window comprised between $1$ and $4$,
while the former property means
that
Monte Carlo studies within this window are feasible.

\subsubsection*{\addcontentsline{toc}{subsubsection}{The single-junction-switch vs. -flip dynamics}The single-junction-switch vs. -flip dynamics}

Single-junction-flip Metropolis generation series came also,
by applying the same annealing schedule scheme,
to equilibria.
Nevertheless,
the single-junction-flip Metropolis generation series contrast with
the single-junction-switch Metropolis generation series
into two major ways:
\begin{enumerate}[i)]
	\item The total amounts of surprisal at single-junction-flip equilibria
		within the real-world window $2$--$3$ \citep{MEJNewmanNI}
		appear to be
		greater than the total amounts of surprisal of the self-fit outputs
		by about $300$ times their standard-deviations
		---
		while the corresponding ones at single-junction-switch equilibria
		are greater
		by at most $8$ times their standard-deviations.
		This makes the single-junction-flip Metropolis algorithm
		clearly inconsistent
		with the self-fit join principle,
		hence unrealistic.
	\item For large effective scaling exponent $\widetilde{\lambda}$ values,
		the single-junction-flip simulations appear much less subject to noise.
\end{enumerate}
In brief:
restricting matchings to maximal matchings
renders our system realistic
but numerically unstable for relatively large scaling exponents;
vice versa,
loosing matchings
renders our system unrealistic
but numerically stable
for a relatively wider range of scaling exponents.

To explain this,
we must keep in mind
that
our particular working assumptions neglect junctions.
In fact,
in one hand,
the single-junction-switch dynamics provides
to our working assumptions a \textit{``hard-coded''} constraint on junction layouts
so that our system becomes more realistic.
On the other hand,
the single-junction-flip dynamics
allows the Metropolis algorithm to reject maximal matchings
in favour of non-maximal matchings
so that our algorithm becomes numerically more stable.
To resolve this dilemma,
we may replace
the hard constraint on junction layouts
with
a soft constraint.
This may take,
in the total amount of surprisal $S_{\mu}$ \eqref{OEMSSOUSN/eq/SOUSN/workinghypotesis/surprisal/amount/total},
the form of additional surprisal terms involving junctions or mixing streets and junctions.
The derivation of such terms is however
outside the scope of the present paper.

\subsection*{\addcontentsline{toc}{subsection}{Unorthodox Watts-Strogatz phase diagram}\label{sec/metropolis/phasediagram}Unorthodox Watts-Strogatz phase diagram}

So far we have demonstrated that Central London sustains
scale-free configurations of streets
over a wide range of scalings.
This means that
the selection
of a realistic scale-free configuration of street
involves other criteria
than just scale-freeness.
An appealing explanation might hold with the small-world crossover,
which may happen as we exposed
in our ``\nameref{sec/modelization/dynamics/variant/comparison}'' section.
This hypothesis illustrates well
the new class of explorations that
the method presented in the present paper
brings in the field.
We keep our hypothesis for future work.
Meanwhile,
to emphasize our contribution,
we demonstrate the small-world crossover
by adopting
the phase diagram used for
Watts-Strogatz small-world models \citep{WattsStrogatz1998,MEJNewmanSFCN2003,MEJNewmanNI}.

\subsubsection*{\addcontentsline{toc}{subsubsection}{Two-regime phase diagram}\label{sec/metropolis/phasediagram/tworegimephase}Two-regime phase diagram}

\begin{figure}[hbp]
	\begin{center}
		\includegraphics[width=0.95\linewidth]{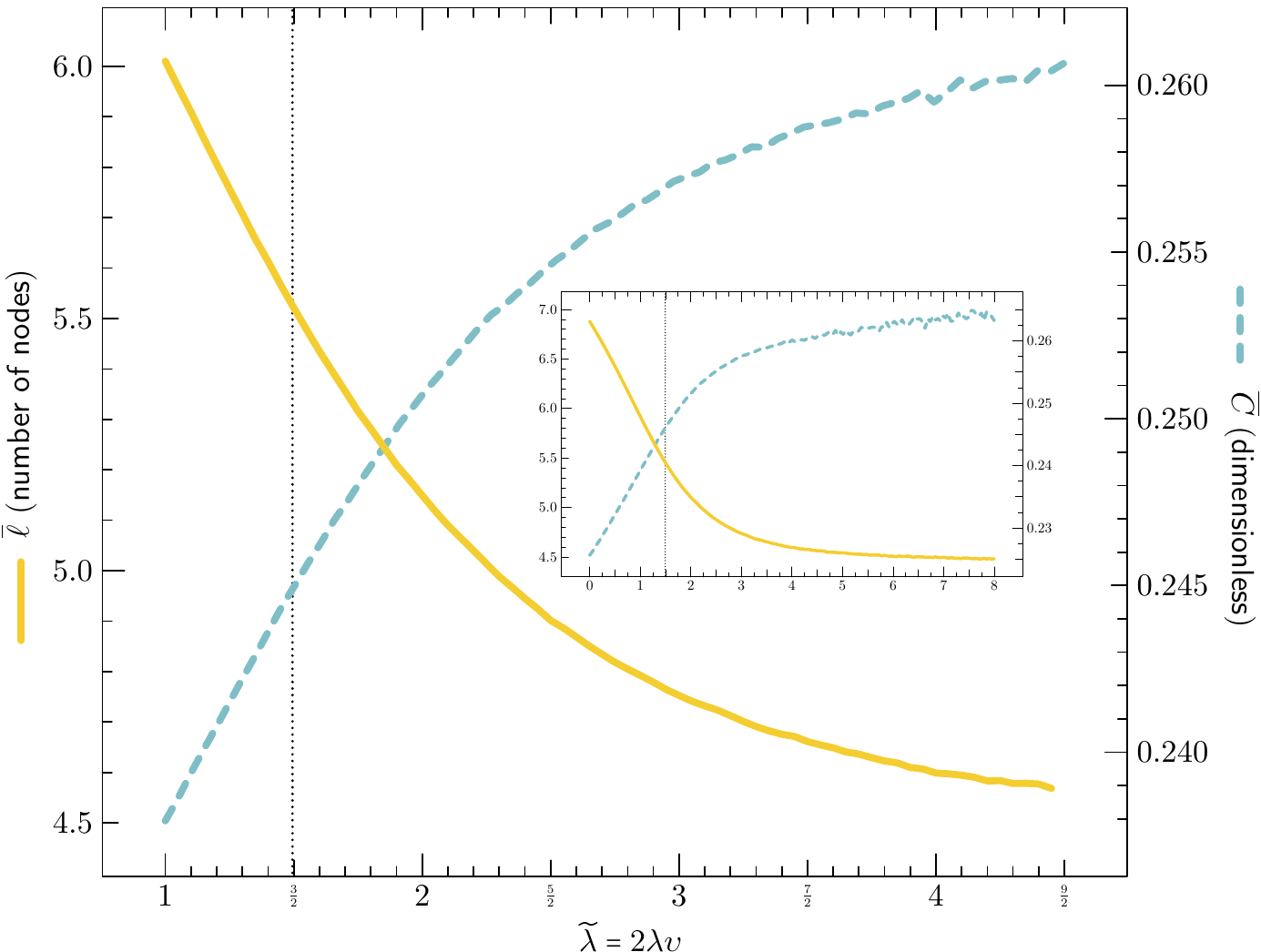}
	\end{center}
	\caption{\label{OEMSSOUSN/fig/surprisal/Metropolis/equilibria/CentralLondon/crossover}%
		Mean geodesic distance between nodes
		and
		mean local transitivity
		versus effective scaling exponent
		for
		the information networks
		of Central London (United Kingdom):
		yellow solid line plots the mean of
		the mean geodesic distance between node pairs
		$\ell$;
		turquoise dashed line plots the mean of
		the mean local transitivity
		$C$;
		the vertical dotted line represents
		the effective scaling exponent $\widetilde{\lambda}_{0}$
		for which the associated Metropolis equilibrium
		approaches on average self-fit outputs
		---
		as illustrated in Figure~{\ref{OEMSSOUSN/fig/surprisal/Metropolis/equilibria/CentralLondon/series}d};
		the inset shows their asymptotic behaviours;
		the main figure and the inset have the same aspect ratio.
		The experimental protocol was the same as in Figure~\ref{OEMSSOUSN/fig/surprisal/Metropolis/equilibria/CentralLondon/behaviour}.
		(%
		We attribute the noise to the same reasons as in Figure~\ref{OEMSSOUSN/fig/surprisal/Metropolis/equilibria/CentralLondon/behaviour}.%
		)%
		}
\end{figure}

The Watts-Strogatz phase diagram for Central London
plotted in Figure~\ref{OEMSSOUSN/fig/surprisal/Metropolis/equilibria/CentralLondon/crossover}
shows two crossovers which occurs simultaneously at the effective scaling value of $3$.
This phase diagram plots
for the information networks of Central London
the averages of
their mean geodesic distance
(or \emph{mean
vertex-vertex distance})
$\ell$
and of
their mean local transitivity
(or \emph{clustering coefficient})
$C$
as functions of the rewiring parameter \citep{WattsStrogatz1998,MEJNewmanNI}.
The rewiring parameter is here
the effective scaling exponent $\widetilde{\lambda}$.
These two functions experience a qualitative change of behaviour
in the vicinity of effective scaling ${\widetilde{\lambda}=3}$,
that is,
they exhibit a crossover at effective scaling ${\widetilde{\lambda}=3}$
\citep{SGluzmanVIYukalov1998}.
Our claim that
the crossovers precisely happen at effective scaling ${\widetilde{\lambda}=3}$
relies on the arguments given by \citet{RCohenSHavlin2003}.
The two simultaneous crossovers indicate two distinct phases or regimes:
\begin{enumerate}[i)]
	\item A uniform regime takes place as effective scaling increases from $3$.
		As effective scaling gets higher and higher starting from $3$,
		the mean geodesic distance between node pairs $\ell$
		(resp. the mean local transitivity $C$)
		tends on average asymptotically towards
		a slightly-decreasing
		(resp. a slightly-increasing)
		plateau.
		The asymptotic behaviours become obvious around the effective scaling value of $4$.
		This means that in this regime the involving phenomena are saturating.
	\item An emergent/reduction regime occurs as effective scaling decreases from $3$.
		As effective scaling gets lower and lower starting from $3$,
		the mean geodesic distance between node pairs $\ell$
		(resp. the mean local transitivity $C$)
		increases
		(resp. decreases)
		on average
		to reach a linear behaviour around the effective scaling value of $3/2$.
		The decreasing on average of the mean geodesic distance between node pairs $\ell$
		as scaling increases
		means that scaling is inducing a smaller world \citep{MEJNewmanNI}.
		The increasing on average of the mean local transitivity $C$
		as scaling increases
		means that scaling is inducing a denser world
		or a world with less \textit{``structural holes''}
		\citep{MEJNewmanNI}.
		In our context,
		a smaller world means
		journeys with lesser changes of streets
		(%
		see
		end of point~\ref{pt/modelization/dynamics/variant/comparison/crossover}
		in ``\nameref{sec/modelization/dynamics/variant/comparison}'' section);
		a denser world means more local alternative routes.
\end{enumerate}
We may regard the two linear behaviours for small and large scalings as degenerate or extreme.
In this sense the relevant part of the phase diagram yields between the effective scaling values of $3/2$ and $4$.
This is consistent with our previous expectation
in ``\nameref{sec/metropolis/equilibria}'' section
that the relevant physics of our system may occur within the window $1$--$4$.

\subsubsection*{\addcontentsline{toc}{subsubsection}{A brief comparison with the classical Watts-Strogatz phase diagram}\label{sec/metropolis/phasediagram/briefcomparison}A brief comparison with the classical Watts-Strogatz phase diagram}

The Watts-Strogatz phase diagram for Central London confirms that
our urban street network model underlies the small-world effect.
Our expectation was sketched
in point~\ref{pt/modelization/dynamics/variant/comparison/crossover}
of our ``\nameref{sec/modelization/dynamics/variant/comparison}'' section.
Nonetheless the obtained Watts-Strogatz phase diagram differs from
the
classical
Watts-Strogatz phase diagram
\citetext{\citealp[Fig.~{6.2}]{MEJNewmanSFCN2003}; \citealp[Fig.~{2}]{WattsStrogatz1998}}
in three essential characteristics:
\begin{enumerate}[i)]
	\item A smaller world means a denser world, not a less dense one.
		This is because on average the mean local transitivity $C$ increases instead of decreasing.
	\item The two crossovers coincide.
		In other words,
		no emergent/reduction regime overlaps with an uniform regime
		and vice versa.
	\item The small-world effect predominates.
		The overall variation of the average of the mean local transitivity $C$ is of order $0.04$,
		so we may regard the local transitivity evolution as insignificant.
		Meanwhile the average of the mean geodesic distance between node pairs $\ell$
		gains overall $2.5$ nodes and $0.9$ nodes within the relevant window from $3/2$ to $4$,
		that is,
		the small-world effect is actually the substantial phenomenon.
\end{enumerate}
Therefore,
contrary to Watts-Strogatz small-world networks \citep{WattsStrogatz1998,MEJNewmanSFCN2003,MEJNewmanNI},
the information networks of Central London
experience no balance between local transitivity and the small-world effect.
Actually,
among both,
only the small-world effect
is relevant as scaling varies.

\subsubsection*{\addcontentsline{toc}{subsubsection}{The self-fit configurations of streets are inefficient}\label{sec/metropolis/phasediagram/selfitinefficiency}The self-fit configurations of streets are inefficient}

Generation series (d) in Figure~\ref{OEMSSOUSN/fig/surprisal/Metropolis/equilibria/CentralLondon/series} shows that,
for Central London,
self-fit outputs are hardly distinguishable from
Metropolis in-equilibrium generations at effective scaling $\widetilde{\lambda}_{0}=1.495$.
This value can be regarded as
a measurement of the effective scaling
at which
Central London
sustains self-fit configurations of streets.
This measurement is represented
on the Watts-Strogatz phase diagram
for Central London
in Figure~\ref{OEMSSOUSN/fig/surprisal/Metropolis/equilibria/CentralLondon/crossover}
by the vertical dotted line.
The phase diagram immediately tells us on self-fit information networks
for Central London
three noteworthy facts:
\begin{enumerate}[i)]
	\item Their worlds are on average of a magnitude one node larger.
		As natural reference,
		we take
		here
		the high-scaling
		asymptotic
		configurations of streets.
	\item They occur around the end of the low-scaling linear behaviour.
		That is,
		they occur around the low-scaling boundary
		of the relevant window $3/2$--$4$.
	\item There is room for information networks with significantly smaller worlds.
		A quick check shows that information network worlds
		at effective scaling $2.50$
		(centre of realistic window $2$--$3$)
		and $2.75$
		(centre of the relevant window $3/2$--$4$)
		are on average,
		respectively,
		$0.6$ and $0.7$ nodes smaller
		than the self-fit worlds.
		These offset drops are,
		respectively,
		of the order of $60\%$ and $70\%$.
		Namely,
		they are substantial.
\end{enumerate}
To summarize,
the Watts-Strogatz phase diagram describes
self-fit information networks
for Central London
as being
on average
relatively large worlds.

However,
Central London dwellers may rather want to know whether
their self-fit configurations of streets are efficient.
Efficiency means
here
for city-dwellers that they can complete their journeys as fast as possible.
This can be partially achieved
by decreasing
as much as possible
the number of street changes required per journey.
This means to develop
information network whose worlds
are as small as possible.
This corresponds
on the Watts-Strogatz phase diagram
to information networks having on average relatively small worlds.
In effect,
this involves the information networks that actually experience the crossover.
As we have seen,
quite the opposite actually happens
to the self-fit information networks of Central London:
they take place where the small-scaling linear behaviour ceases
and
they underlie on average relatively large worlds.
In brief,
the self-fit configurations of streets for Central London are inefficient.

\section*{\addcontentsline{toc}{section}{Conclusions and future works}\label{sec/conclusions}Conclusions and future works}

Unplanned or self-organized
urban street networks undergo a scale-free coherence
that we interpret
in terms of a fluctuating system.
This paper sketches
how the Metropolis algorithm,
which embodies well the idea of fluctuating systems \citep{MCMSP,DPLandauKBinderGMCSSP},
can apply
to self-organized urban street networks
once our interpretation is embraced.
The Metropolis algorithm is a classical entry-point for more elaborate Monte Carlo methods.
These methods are the natural numerical companions for theoretical studies on fluctuating systems,
and vice versa.
Our theoretical framework is the maximum entropy formalism
(\textsc{MaxEnt})
\citep{WTGrandyJrFSMET,ETJaynesPTLS,ALawrencePP2019}.

Our prior hypothesis \citep{WTGrandyJrFSMET,ETJaynesPTLS}
is scale-freeness \citep{HEStanleyIPTCP}.
Assuming this property as
the result of an underlying self-similarity symmetry \citep{BMandelbrot1982,MBatty2008}
paves the way to
a symmetry-conservation correspondence as used in physics \citep{DJGrossTRSFP1996,DRomeroMaltrana2015}.
This physical idea effortlessly adapts itself to \textsc{MaxEnt}.
This allows us,
as required for implementing any Monte Carlo method,
to set up our prescribed statistical equilibrium.
The self-similarity symmetry demands the conservation on average of the logarithm of an extensive quantity
which,
by virtue of \textsc{MaxEnt},
most plausibly underlies a discrete Pareto distribution \citep{AClausetCRShaliziMEJNewman2009}.
The scaling exponent is our equilibrium parameter.
Meanwhile,
the best we can tell on any information network is that
it is a mesoscopic system whose objects,
nodes (streets) and edges (junctions),
have equiprobable configurations.
So,
the best we can assume about our extensive quantity is that
it is a number of equiprobable configurations.
The conserved quantity
becomes then an average of Boltzmann entropies.
However
we may rather interpret this information measure
as an amount of surprisal \citep{MTribusTT,DJCMacKayITILA,DApplebaumPIIA,JVStone2015,ALawrencePP2019} that
actually
quantifies the comprehension of the city-dwellers for their own urban street network \citep{JBenoitOPSOUSN2019,SESOPLUSN}.
Once our prescribed statistical equilibrium is fully set up,
we can readily implement our Metropolis acceptance ratio.

As concerns the ergodic dynamics,
its counterpart,
the nonoverlapping walk approaches found in the literature
\citep{PortaTNAUSDA2006,BJiangSZhaoJYin2008,MPViana2013}
appear inappropriate but nonetheless inspirational.
We imagine information networks
not in terms of haphazard nonoverlapping walks along street-segments,
but in term of random street layout at junctions.
Our approach readily leads to dynamics that mimic
the classical single-spin-flip dynamics in Ising models \citep{MCMSP,DPLandauKBinderGMCSSP,JBerlinskyABHarrisGTPSM}.
At every junction,
each pair of street-segments that can hold a street is a link of a graph where street-segments map to nodes,
so that
each matching \citep{PEMMARJU} of this graph
represents a possible layout.
As the single-spin-flip dynamics changes the state of a spin into another possible state,
our dynamics changes the matching (layout) of a junction into another possible matching (layout).
We named
single-junction-flip the dynamics that involves any matchings,
and
single-junction-switch the dynamics that involves only maximal matchings \citep{PEMMARJU}.
If our approach implicitly implies that
self-organized urban street networks might sustain scaling coherence over a wide range of scalings,
finding dynamics reminiscent of Ising models suggests first and foremost that
they might undergo a crossover as scaling varies.
Since large scale-free networks exhibit ultra-small- and small-world behaviours
for scaling values respectively smaller and greater that $3$ \citep{RCohenSHavlin2003},
self-organized urban street networks might actually experience
as scaling increases
a small-world crossover
around the scaling value of $3$.

We choose as case study
the recognized self-organized urban street network of Central London (United Kingdom) \citep{GreatStreets1993}.
Simulations based on
predominant streets
and
an asymptotic agent model driven by social interactions \citep{YDover2004,JBenoitOPSOUSN2019,SESOPLUSN}
show that
the single-junction-switch Metropolis algorithm generates equilibria that are consistent
with the aforementioned nonoverlapping walk approaches.
The simulations remain consistent
over a range of scaling exponents
large enough to contain the realistic window from $2$ to $3$ \citep{MEJNewmanNI}
and to capture changes of behaviour
in their total and average amounts of surprisal.
Thusly,
the single-junction-switch Metropolis algorithm allows simulational investigations.
The single-junction-flip dynamics also leads to equilibria,
but with unrealistic amounts of surprisal.
We explain this,
given that
our model neglects junctions
while
the single-junction-switch dynamics coerces junctions to have maximal layouts,
by a lack of constraints on junctions.
Along this explanation,
the single-junction-flip dynamics may allow to investigate
the role played by junctions.
In brief,
our simulations on Central London
show
that our adaptation of the Metropolis algorithm
for generating self-organized information networks
is applicable and relevant.

To illustrate our
innovative methodology,
we plot the Watts-Strogatz phase diagram
with scaling as rewiring parameter.
The phase diagram exhibits
an emergent/reduction regime followed by an uniform regime
as scaling increases.
That is,
the small-world and the local transitivity crossovers occur simultaneously.
However
only the former is significant in magnitude.
The crossovers happen approximately around the scaling value of $3$.
More noticeably,
the crossovers curve within the realistic window from $2$ to $3$.
Thusly,
as expected,
our phase diagram demonstrates a small-world crossover around the scaling value of $3$.
Our phase diagram also allows us to discuss the pertinence of the self-fit outputs.
The self-fit outputs take place on average nearly the scaling value of $3/2$,
namely,
significantly before the realistic window $2$--$3$.
They actually occur on average at the ending of the linear scaling behaviour observed at low scalings
---
which we may consider as degenerate.
Concretely
this means that self-fit outputs generate on average information networks that underlie relatively large worlds,
namely,
that are inefficient.
If the implicit belief that
self-organized urban street networks have reached an optimal balance over time
holds,
representative information networks
may rather occur within the realistic window $2$--$3$
where their worlds are on average relatively small,
namely, efficient
---
assuming that the small-world effect gets counterbalanced as its effect curves.
Thusly,
our
illustrative
Watts-Strogatz phase diagram
challenges the state-of-the-art on generating information networks,
while
it
indicates that
self-organized information networks can undergo
as scaling increases
a small-world crossover
curving within the realistic window $2$--$3$.
In other words,
our illustrative numerical \textit{``experiment''}
on Central London
demonstrates that
our adaptation of the Metropolis algorithm
for generating self-organized information networks
is
indeed
pertinent to gain new insights.

From a fundamental point of view,
future works must focus on two points.
First,
we must recognize the deep origin underlying
the extensive quantity associated to the scaling exponent
in order to specify its very nature.
Second,
we must find an uncoercive way to involve junctions
in order to investigate their role.
From a simulational point of view,
we must investigate the undergoing small-world crossover
by considering
other network phenomena \citep{MEJNewmanNI}
and
a large panel of recognized self-organized urban street networks \citep{GreatStreets1993,CrucittiCMSNUS2006}.
We anticipate to observe network phenomena \citep{MEJNewmanNI}
that counterbalance
the small-world effect
within
or around
the realistic window $2$--$3$ \citep{MEJNewmanNI}.
From an observational point of view,
our fluctuating approach clearly challenges the current method
to determine the scaling exponent of an urban street network
which is based on a single arbitrary output \citep{PortaTNAUSDA2006,BJiangSZhaoJYin2008}.
In view to confront our simulational data against observational data,
we must derive methods able
to \textit{``take''} the scaling exponent
and
to measure
network measures \citep{MEJNewmanNI}
along with their
susceptibilities \citep{MCMSP,WTGrandyJrFSMET,ETJaynesPTLS}.
From a practical point of view,
we envision that our
Metropolis adaptation may initiate,
alongside Monte Carlo models getting more elaborate but also more realistic,
a
\textit{`scaling-dynamics'} based
description
of our urban street networks and,
by extension,
of our cities.
Such a descriptive framework
may provide
fruitful analogies
with thermodynamics
and
precious insights
on unplanned evolution
for
city scientists,
city designers,
and
decision-makers
to anticipate
the evolution
of our cities.

\printendnotes*[OEMSSOUSN]

\begin{additionalfiles}[10.5281/zenodo.4439354]
	\item\label{OEMSSOUSN/ani/streetsegments/joinedsequence/notionalexamples}
		State-of-the-art construction paradigm for configurations of streets (Animation A1)%
		\embedadditionalfile[mimetype={application/pdf},desc={Animation A1}]{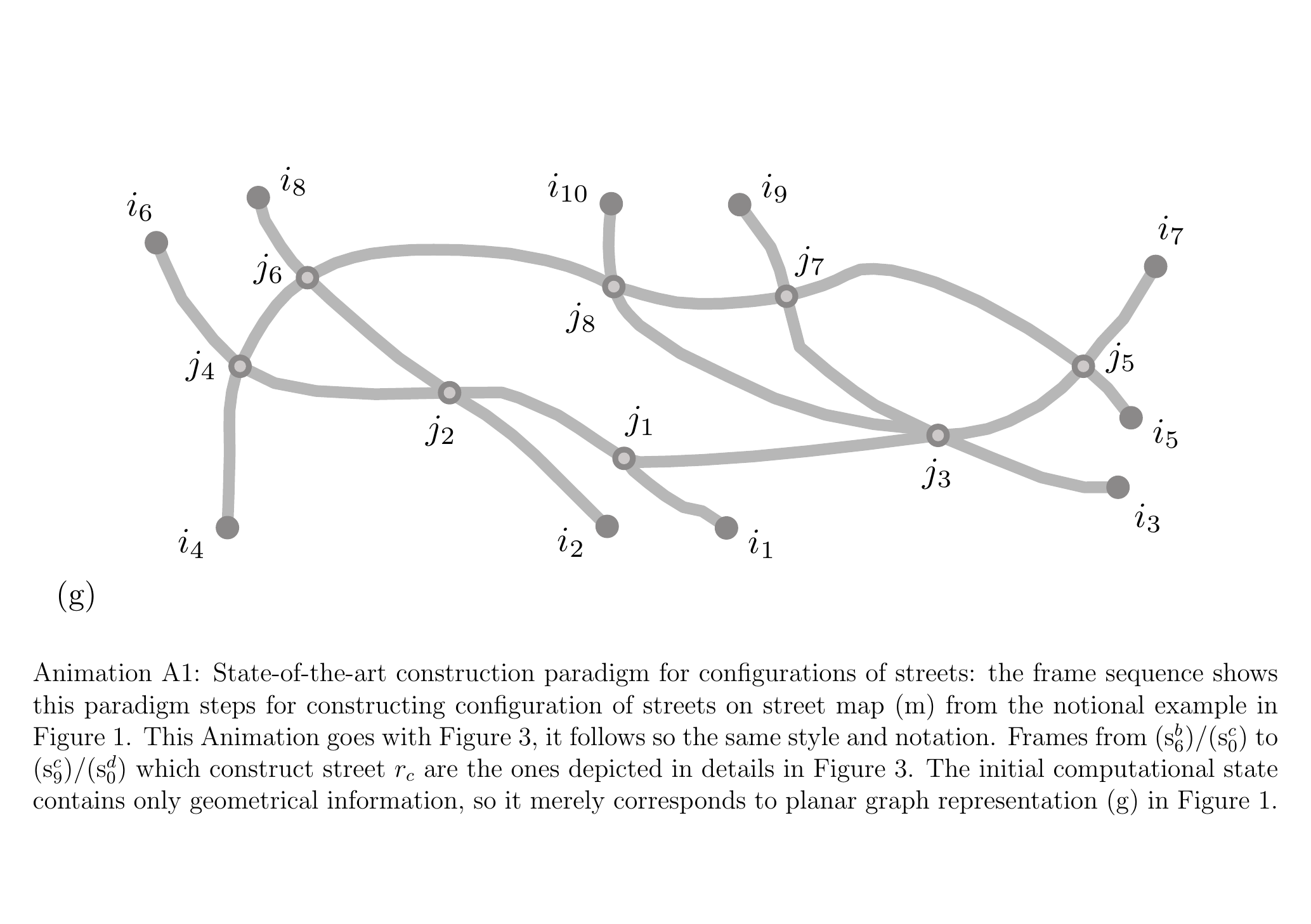}
	\item\label{OEMSSOUSN/ani/junctions/flipflapflopsequence/notionalexamples}
		Single-junction-switch Metropolis algorithm for configurations of streets (Animation A2)%
		\embedadditionalfile[mimetype={application/pdf},desc={Animation A2}]{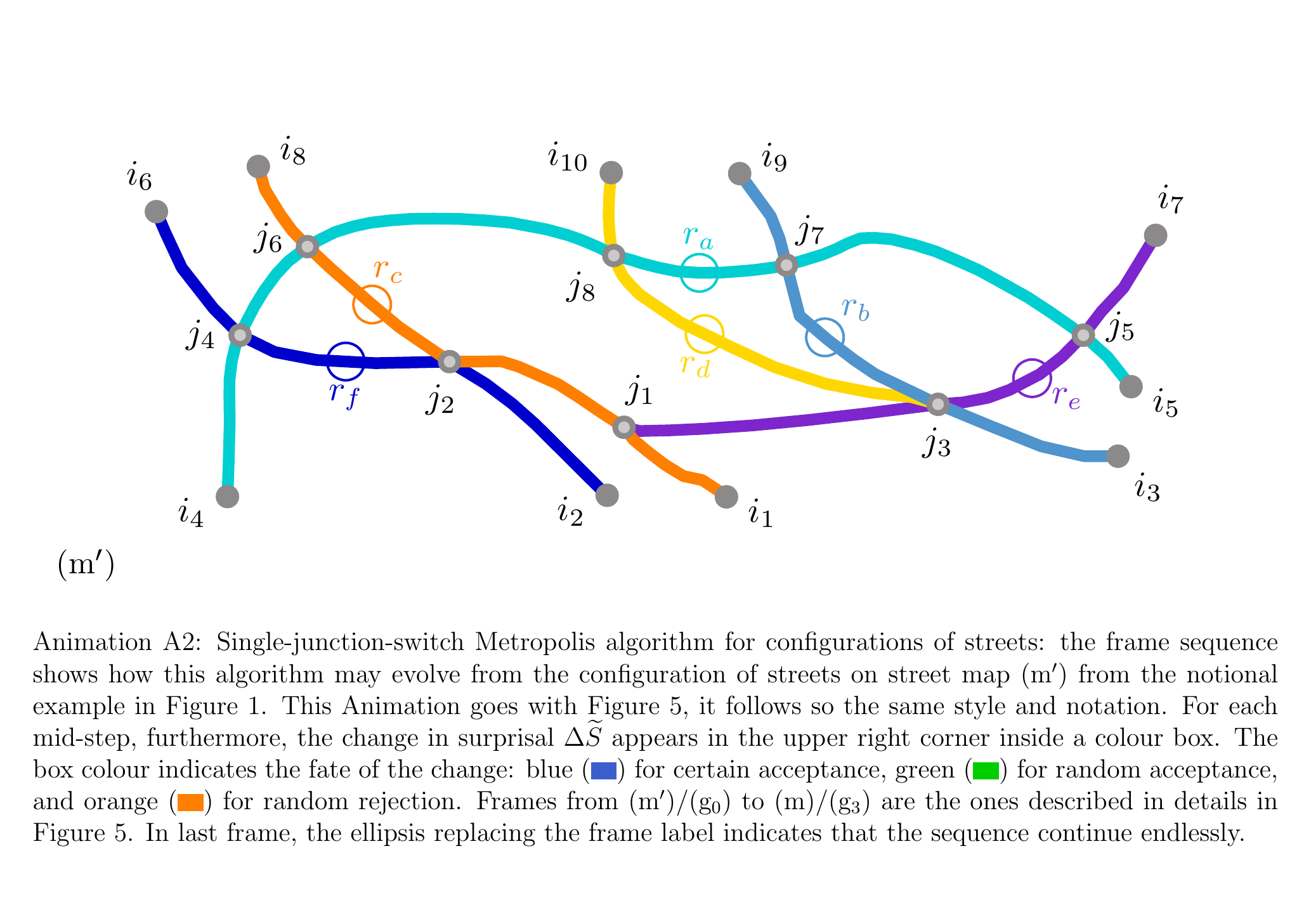}
\end{additionalfiles}

\begin{backmatter}
\section*{Abbreviations}

\textsc{MaxEnt}:
	Maximum Entropy formalism (or principle);
\textsc{SE}: Scaling Equilibrium;
\textsc{TE}: Thermal Equilibrium.

\section*{\label{bm/sec/availability}Availability of data and materials}

The map of the urban street network of Central London was extracted
from the Open Street Map (\textsc{OSM}) comprehensive archive \citep{OpenStreetMapAdHoc}
and simplified
with \texttt{OSMnx}~${\mathrm{v}0.11.4}$ \citep{OSMnxGBoeing2017}.
The network measures were computed with the \texttt{igraph} \texttt{C} library ${\mathrm{v}0.8.2}$ \citep{GCsardiTNepusz2006,IGRAPH000802}.
The software used to perform the simulations is available,
along with the map and generation series samples,
at \onlinedoi{10.5281/zenodo.3746140}.
The datasets generated and analysed during the current study are available
from the corresponding author on reasonable request.

\section*{Author's contributions}

{JGMB}
	conceived and designed the study,
	collected and treated the map data,
	designed and programmed the simulation tools,
	performed and treated the simulations,
	and
	wrote the manuscript.
{SEGJ}
	helped to shape the manuscript.
Both authors read and approved the final manuscript.

\section*{Acknowledgements}

This work was supported by
the NYUAD Center for Interacting Urban Networks (CITIES),
funded
by Tamkeen under the NYUAD Research Institute Award CG001
and
by the Swiss Re Institute
under the Quantum Cities\texttrademark{} initiative.

\section*{Competing interests}

The authors declare that they have no competing interests.


\end{backmatter}

\end{document}